\documentclass[superscriptaddress,showpacs,aps,twocolumn,prb,floatfix]{revtex4-2}
\usepackage{eurosym}
\usepackage{amssymb}
\usepackage{amsmath}
\usepackage{amsfonts}
\usepackage{amsbsy}
\usepackage{graphicx}
\usepackage{bm}
\usepackage{xcolor}
\usepackage{hyperref}
\usepackage{units}

\usepackage[normalem]{ulem}

\begin{document}

\title{Iron phthalocyanine on Au(111) is a ``non-Landau'' Fermi liquid}

\author{R. \v{Z}itko}
\affiliation{Jo\v{z}ef  Stefan  Institute,  Jamova  39,  SI-1000  Ljubljana,  Slovenia}
\affiliation{Faculty  of  Mathematics  and  Physics,  University  of  Ljubljana,  Jadranska  19,  SI-1000  Ljubljana,  Slovenia}
\email{rok.zitko@ijs.si}

\author{G. G. Blesio}
\affiliation{Jo\v{z}ef  Stefan  Institute,  Jamova  39,  SI-1000  Ljubljana,  Slovenia}
\affiliation{Instituto de F\'{\i}sica Rosario (CONICET) and Universidad Nacional de Rosario, 
Bv. 27 de Febrero 210 bis, 2000 Rosario, Argentina}

\author{L. O. Manuel}
\affiliation{Instituto de F\'{\i}sica Rosario (CONICET) and Universidad Nacional de Rosario, 
Bv. 27 de Febrero 210 bis, 2000 Rosario, Argentina}

\author{A. A. Aligia}
\affiliation{Instituto de Nanociencia y Nanotecnolog\'{\i}a CNEA-CONICET,
Centro At\'{o}mico Bariloche and Instituto Balseiro, 8400 Bariloche, Argentina}

\begin{abstract}
\end{abstract}

\maketitle

\section*{Abstract}

The paradigm of Landau's Fermi liquid theory has been challenged with
the finding of a strongly interacting Fermi liquid that cannot be
adiabatically connected to a non-interacting system. A spin-1
two-channel Kondo impurity with anisotropy $D$ has a quantum phase
transition between two topologically different Fermi liquids with a
peak (dip) in the Fermi level for $D < D_c$ ($D > D_c$).

Extending this theory to general multi-orbital problems with finite
magnetic field, we reinterpret in a unified and consistent fashion
several experimental studies of iron phthalocyanine molecules on
Au(111) that were previously described in disconnected and conflicting
ways. The differential conductance shows a zero-bias dip that widens
when the molecule is lifted from the surface (reducing the Kondo
couplings) and is transformed continuously into a peak under an
applied magnetic field. We reproduce all features and propose an
experiment to induce the topological transition.

\section*{Introduction}
%
%
Transport properties of single molecules in contact with metal electrodes 
are being extensively studied due to their potential use as active components of new electronic 
devices \cite{naturefocus,cuevas,evers20}. The paradigmatic Kondo effect 
(screening of molecule's magnetic moment by conduction band electrons from metal) 
is ubiquitous in these systems and allows to control the current by different 
external parameters \cite{liang02,yu05,leuenberger06,oso1,parks07,roch08,roch09,oso2,parks10,florens11,
vincent12,mina,hein,orma,hira,yang}.

The Kondo model (KM) for a molecule with spin $S$ coupled to $n$ conduction bands with different 
symmetries (channels) can be classified into three types: underscreened KM when the number of channels is 
too small to fully screen the spin, overscreened KM the other way, and compensated Kondo model (CKM) at the frontier. 
If there is only one way to build the spin, the frontier is at $n = 2S$ \cite{nozieres80}.
The ground state of an underscreened molecule has a non-zero residual spin, and is a singular Fermi liquid with logarithmic
corrections at low energies~\cite{mehta05,logan09}. In particular, the case
with $S=n=1$ and residual spin $1/2$ has been 
thoroughly investigated in
molecular nanoscopic systems~\cite{parks07,roch09,florens11,logan09,cornaglia11}. The overscreened
model is a non-Fermi liquid that exhibits even more striking singular properties at low
temperatures~\cite{oreg,ifti,zhu}. 
In general it was, however, strongly believed that
the CKM always corresponds to an
ordinary Fermi liquid (OFL) with regular low-energy behaviour showing no non-analytical 
features. Nevertheless, it has been recently shown that this changes in the presence of an
anisotropy term $DS_z^2$ (Refs. \onlinecite{blesio18,blesio19}), which can drastically modify the
ground state \cite{otte08,zitko08,zitko10,dinapoli,oberg13}. 

In a Fermi liquid, the life-time of the quasiparticles with excitation energy $\omega$ scales as 
$\omega^{-2}$ close to the Fermi level at zero temperature, due to restrictions of phase space
imposed by the Pauli principle, as first shown by L. Landau. 
This picture applies to all weakly interacting metals, but it may be broken in the presence of strong interactions. 
In particular, in the overscreened KM the quasiparticles are not well-defined even at the Fermi level, 
and in the underscreened KM they have a logarithmic dependence on $\omega$ for small 
$\omega$~\cite{mehta05,logan09}. 
The CKM with $D=0$ (and the more general Anderson impurity model from which the CKM is derived) corresponds 
to an OFL. In this case, the Friedel sum rule relates the zero-temperature spectral density of the localized 
states at the Fermi level ($\omega=0$), as well as the zero-bias conductance, with the ground-state occupancy 
of these states \cite{lang,yoshi}. As a consequence of the fractional occupancies per channel and spin in the 
Kondo regime, the spectral density has a Kondo peak at the Fermi level and the differential conductance $dI/dV$
a zero-bias anomaly, i.e., a peak at voltage $V=0$. 

However, for the $S=1$ CKM, this picture changes dramatically for $D > D_c$, where $D_c$ is the critical 
anisotropy at which a topological quantum phase transition takes place. 
The spectral density at $\omega=0$  
vanishes. 
In this case, the Friedel sum rule should be modified to allow 
for a non-zero value of a Luttinger integral $I_L$, which has a topological 
character with a discrete set of possible values. A system in this regime is
a Fermi liquid, yet it
cannot be adiabatically connected to a non-interacting system. 
A closer scrutiny reveals the presence of a $\delta$-peak exactly at the
Fermi level in the imaginary part of the impurity self-energy, 
even though the low-energy
scattering properties are not in any way anomalous and remain
Fermi-liquid like.
Such a ground state has been called a non-Landau Fermi liquid (NLFL) \cite{blesio18,blesio19}. 
Previously, other strongly
interacting models have been shown to display NLFL behavior for some 
parameters~\cite{curtin18,nishi}.

A set of recent spectroscopic measurements performed on iron phthalocyanine (FePc) on Au(111)
showed features reminiscent of those expected in NLFL systems.
Two
relevant experiments, performed  before the theory for the NLFL
system was developed \cite{mina,hira}, reported basic characteristics
of the system. Another more recent work
\cite{yang} is a particularly revealing detailed
study of the magnetic field and temperature dependence. The existing
interpretations proposed in these three works have serious
deficiencies, as we discuss in the following. The
NLFL theory accounts for the totality of
the available results on FePc/Au(111), as well as for very recent
experiments on MnPc/Au(111) \cite{guo}.
A non-trivial extension of the existing theory
to nonequivalent channels and finite magnetic fields 
presented in this work, was necessary to make these inferences.
We also explain how to observe experimentally the topological
quantum phase transition between a non-Landau and a Landau Fermi
liquid.

Ab initio calculations \cite{mina} show that the molecule in the ``on top'' position 
has a spin $S=1$ formed by an electron in the 3d orbital of Fe with symmetry $3z^2-r^2$ 
and another shared between the degenerate $\pi$ ($xz$, $yz$) Fe 3d orbitals. 
The former has larger hybridization with the Au substrate than the latter.
The molecule can be described
by a three-channel multi-orbital Anderson Hamiltonian, 
which is compensated \cite{joaq}. 
The low-temperature differential conductance $dI/dV$ measured by a 
scanning-tunneling microscope (STM)
at zero magnetic field shows a broad peak of half-width 
$\sim 240$ K \cite{mina,joaq} centered near 
$V=0$ and superposed to it a narrow dip of half width $\sim 2.7$ K \cite{hira},
also centered near $V=0$.
This is exactly the shape expected in the NLFL regime of the anisotropic $S=1$ CKM near the 
topological quantum phase transition at $D=D_c$ \cite{blesio18,blesio19}.

In the absence of these theoretical results, the first set of experiments  \cite{mina} 
was initially interpreted as a two-stage Kondo effect, with the larger energy 
scale corresponding to the stronger hybridized channel with $3z^2-r^2$ symmetry,
and the smaller one due to the $\pi$ orbitals \cite{mina,joaq}. The
magnetic anisotropy was not taken into account.
In general, as a consequence of interference effects, a $dI/dV$ curve has a peak when the tip 
of the STM 
has a stronger hopping to the localized orbitals and a dip when the tip 
has instead a stronger hopping to the conduction electrons \cite{uj,remir} Therefore, this 
interpretation requires that the tip has a stronger (weaker) hybridization to the 
localized $3z^2-r^2$ ($\pi$) orbitals than to the conduction electrons of the same 
symmetry in addition to appropriate magnitudes of these hybridizations \cite{joaq}.
This seems unlikely.

The more recent experiment in which the molecule is raised from the surface, 
thereby reducing the hybridization of both localized states to the Au substrate,
provides a stringent test of the above interpretation \cite{hira}. In the original
interpretation based on the two-stage Kondo model, both the peak and the dip should 
narrow and become steeper because both Kondo temperatures decrease. Instead,
the experiments show clearly that the dip broadens and
the peak flattens as the molecule is raised, in agreement with the expectations 
for the NLFL (see below). The theoretical explanation of the authors, further discussed in Note 6 of the supplemental
material of their paper.
is only qualitative and considers two contributions with many parameters.

The study of magnetic field effects \cite{yang} unveils a surprising
transformation of the narrow dip
into a peak with increasing field strength $B$. 
This is a challenge to all existing theories so far.
Based on ab initio calculations this effect has been interpreted as a rearrangement 
of the presence of localized orbitals of $3z^2-r^2$ and $\pi$ symmetry near the Fermi 
level \cite{yang}. However, the difference between the energies of these states, of the order 
of 1 eV \cite{mina}, is much larger than the Zeeman energy of about 10T. 
Furthermore, many-body effects, which play a crucial role by shifting the relative positions of peaks in the density of states~
\cite{feiguin} and in the Kondo effect \cite{hewson97} 
have been altogether neglected. Therefore this explanation is very unlikely.

A similar effect of $B$ has been observed in MnPc on Au(111)
which has a similar occupancy of the 3d states as FePc. 
The authors of that work proposed an interpretation in terms of a quantum phase transition 
involving localized singlet states \cite{guo}. This transition, first discussed in the context 
of Tm impurities \cite{allub} and more recently in quantum dots \cite{hof,paas}, in particular 
with C$_{60}$ molecules \cite{roch08,florens11,roura09,roura10}, might be
qualitatively consistent with the experiment (see Fig. 10 of Ref. \onlinecite{roura10}).
However, this requires that the singlet be below the triplet by a few
\unit{meV}, while in fact the triplet is energetically favored by a Hund's coupling of
the order of \unit[1]{eV} \cite{mina}. 

In this work, we show that the behavior of FePc on Au(111) can be
interpreted in a unified and consistent way in terms of an anisotropic
$S=1$ CKM in the parameter regime where the system is a NLFL.
Using the numerical renormalization group (NRG) we solve the anisotropic $S=1$ CKM for
inequivalent channels in the presence of magnetic field and show that all phenomena observed in
the above mentioned experiments can be explained in a consistent and simple way. 
We underpin
these calculations by generalizing the topological theory of the Friedel sum rule to the case in
which channel and spin symmetries are broken. 
In particular, we show how the discontinuous transition at zero magnetic field governs the molecule's excitation spectrum at finite fields, and why the evolution of the spectral line-shape from a dip to a peak with increasing field is nevertheless continuous.

\section*{Results}

\subsection*{Topological quantum phase transition and generalized Friedel sum
rule}

The essence of the atomic multi-orbital three-channel Anderson impurity
model for FePc, which is difficult to handle numerically, is captured
by the $S=1$ two-channel Kondo model with the Hamiltonian
\begin{eqnarray}
H_{K} &=&\sum_{k\tau \sigma }\varepsilon _{k\tau }c_{k\tau \sigma }^{\dagger
}c_{k\tau \sigma }+\sum_{k\tau \sigma \sigma ^{\prime }}\frac{J_{\tau }}{2}
c_{k\tau \sigma }^{\dagger }(\vec{\sigma})_{\sigma \sigma ^{\prime
}}c_{k\tau ^{\prime }\sigma ^{\prime }}\cdot \vec{S}  \notag \\
&&+DS_{z}^{2}-BS_{z},  \label{hk}
\end{eqnarray}
where $c_{k\tau \sigma }^{\dagger }$ creates an electron in the Au substrate
with wave vector $k$, pseudospin $\tau $ (representing a channel with
symmetry $3z^{2}-r^{2}$\ for $\tau =1$ and $\pi $ for $\tau =-1$) and spin $\sigma $. 
The first term describes the substrate bands, the second the Kondo exchange with the localized spin $\vec{S}$ with exchange
couplings $J_{1}>J_{-1}$, the third term is the single-ion
uniaxial magnetic anisotropy,
and the last term is the effect of an applied magnetic field $B$.
This model is equivalent to that used by Hiraoka et al. \cite{hira}
for large Hund's coupling $J_{H}$. As we show below, this model with three
parameters explains all the experimental findings in FePc \cite{note}. Its fundamental ingredients are well 
justified by ab-initio and/or ligand field multiplet calculations: the existence of an almost integer 
electronic occupancy of the Fe ion~\cite{hira,stepanow11}, the spin $S=1$ of the FePc molecule~\cite{mina,hira,yang,mabrouk20}, the markedly 
different hybridizations of the $d_{z^2}$ and $d_\pi$ orbitals with the Au substrate~\cite{mina,hira,kezilebieke14}, and the  
presence of a non-negligible easy-plane magnetic anisotropy. We also 
consider that the gold conduction bands have a constant density of states, as first-principle 
calculations~\cite{gao07} do not find any signature of sharp structures (with widths of the order of $T_K^{(1)}$ or less), 
that would modify sensibly the low energy Kondo physics, and we take the half band width $W=1$ as the unit of energy.

It is worth to stress that, in this work, we are looking for a minimal model that, on one hand, allows 
to understand, in a unified and rather simple way, the low-energy behavior of FePc on Au(111), and that, 
on the other hand, is suitable for a numerically exact resolution at very low energies. 
A more realistic and quantitative modeling for this and other transition metal phthalocyanines 
on metallic surfaces should take into account the full complexity of the 3d orbital manifold and the hybridization with the surface~\cite{mugarza12,jacob13,kugel14,kugel15,valli20}.

The generalized Friedel sum rule is more conveniently discussed in terms of an auxiliary
two-channel Anderson model $H_{A}$ from which $H_{K}$ can be derived in the limit
of total local occupancy pinned to two (one electron in each orbital). This model is
\begin{eqnarray}
H_{A} &=&\sum_{\tau \sigma }\epsilon _{\tau }d_{\tau \sigma }^{\dagger
}d_{\tau \sigma }+\sum_{\tau } U_\tau n_{\tau \uparrow }n_{\tau \downarrow
}+
\notag \\
&-&J_{H}{\vec{S}}_{1}\cdot {\vec{S}}
_{-1}+DS_{z}^{2}-BS_{z}  \notag \\
&+&\sum_{k\tau \sigma }\varepsilon _{k\tau }c_{k\tau \sigma }^{\dagger
}c_{k\tau \sigma }+\sum_{k\tau \sigma }\left( V_{\tau }{c}_{k\tau \sigma
}^{\dagger }{d}_{\tau \sigma }+\mathrm{H.c.}\right) ,  \label{ha}
\end{eqnarray}
where $d_{\tau \sigma }^{\dagger }$ creates a hole with energy $\epsilon
_{\tau }$ in the $d$ orbital $\tau $, $n_{\tau \sigma }=d_{\tau \sigma
}^{\dagger }d_{\tau \sigma }$, and $n_{\tau }=\sum_{\sigma }n_{\tau \sigma }$. 
$\epsilon_\tau$ and $U_\tau$ are the energy level and the Coulomb
repulsion chosen such that $\epsilon_\tau = -U_\tau/2$ to fix the
occupancy in each orbital to one, while the hopping $V_{\tau}$ characterizes the tunneling between the
localized and conduction states with symmetry $\tau$. The Hund's coupling $J_H$
is responsible for the formation of the $S=1$ degree of freedom.
The actual Coulomb interaction contains more terms \cite{blesio18,blesio19,valli20}, 
but they are irrelevant for realistic parameters for FePc.
The two models are
related by the Schrieffer-Wolff transformation, such that $J_\tau
\propto V_\tau^2/U_\tau$ (further details are given in Supplementary Note 1).

The impurity spectral
function per orbital and spin, at the Fermi level and $T=0,$ is related to
the quasiparticle scattering phase shift $\delta _{\tau \sigma }$ by  \cite{lang,taylor,friedel,lutti,langer,shiba} 
\begin{equation}
\rho_{\tau \sigma }(\omega =0)=-\frac{1}{\pi }\mathrm{Im}G^d_{\tau \sigma
}(0)=\frac{1}{\pi \Delta _{\tau }}\sin ^{2}\delta _{\tau \sigma },
\label{rho}
\end{equation}
where $G^d_{\tau \sigma }(\omega )$ is the impurity Green's function $\left\langle
\left\langle d_{\tau \sigma };d_{\tau \sigma }^{\dagger }\right\rangle
\right\rangle $ and $\Delta _{\tau }=\pi \sum_{k}|V_{\tau }|^{2}\delta
(\omega -\varepsilon _{k})$ is the hybridisation strength, assumed independent of energy. 
According to the generalized Friedel sum rule, 
for wide constant unperturbed conduction bands one has 
\begin{equation}
\delta_{\tau \sigma }=\pi \left\langle n_{\tau \sigma }\right\rangle
+I_{\tau \sigma },\text{ }I_{\tau \sigma }=\mathrm{Im}\int_{-\infty
}^{0}d\omega\, G^d_{\tau \sigma }(\omega )\frac{\partial 
\Sigma ^d_{\tau \sigma}(\omega )}{\partial \omega },  \label{del}
\end{equation}
where $\Sigma^d_{\tau \sigma }(\omega )$ is the impurity self energy.
Further details are given in Supplementary Note 2.

Until recently, 
based on the seminal perturbative calculation of Luttinger \cite{lutti}, 
it was generally assumed that the Luttinger
integrals $I_{\tau \sigma }$ vanish (at least for $B=0$). 
However, several cases are now known where $I_{\tau \sigma }=\pm
\pi /2$ \cite{logan09,blesio18,blesio19,curtin18,nishi},
yet the quasiparticle scattering phase shifts show no
low-energy singularities.
In particular in our case for $B=0$
and both orbitals equivalent, the four $I_{\tau \sigma }=I_{L}$ are equal
by symmetry and $I_{L}$ has a topological character, being equal
to $0$ for $D<D_{c}$ and $\pi/2$ for $D>D_{c}$ \cite{blesio18,blesio19}. 
In this case, 
Eqs.~\eqref{rho} and \eqref{del} imply that in the Kondo limit, 
where $\left\langle n_{\tau \sigma
}\right\rangle =1/2$, the spectral densities $\rho _{\tau \sigma }(0)$ jump
from $\rho _{0}=1/(\pi \Delta _{\tau })$ for $D<D_{c}$ to 0 for $D>D_{c}$.
In the general case, when the channels are not equivalent and a magnetic field is present, 
we show in Supplementary Note 2 
that the conservation laws imply that
three topological quantities $T,$ $T_{\tau },$ $T_{\sigma }$ can still be
defined:
\begin{equation}
T=\sum_{\tau \sigma }I_{_{\tau \sigma }},\text{ }T_{\tau }=\sum_{\tau \sigma
}\tau I_{_{\tau \sigma }},\text{ }T_{\sigma }=\sum_{\tau \sigma }\sigma
I_{_{\tau \sigma }},  \label{tn}
\end{equation}
where $\sigma =1$ (-1) for spin $\uparrow $ ($\downarrow $). 
Although previous perturbative calculations assumed $T = T_\tau = T_\sigma \equiv 0$ \cite{yoshi}, 
we find numerically that $T_{\tau }=T_{\sigma} \equiv 0$, but $T=4I_{0}$ with $I_{0}$ 
equal to either $0$ or $\pi/2$. 
This allows us to write the individual Luttinger integrals in the general form
\begin{equation}
I_{1\uparrow }=I_{-1\downarrow }=I_{0}-\alpha,\text{ }I_{1\downarrow
}=I_{-1\uparrow }=I_{0}+\alpha,  \label{il2}
\end{equation}
where $\alpha(D,B)$ is unknown. For $B=0$, by symmetry $I_{_{\tau \uparrow
}}=I_{_{\tau \downarrow }}$ implying $\alpha=0$, 
and therefore not only $T$, $T_\tau$ and $T_\sigma$ but also
individual Luttinger integrals
$I_{\tau\sigma}$ have a topological character. The establishment of
the general form in Eq.~\eqref{il2} and its numerical validation
in the full parameter space of the model is one of the central
results of this work.

\begin{figure}[ht]
\begin{center}
\includegraphics*[width=\columnwidth]{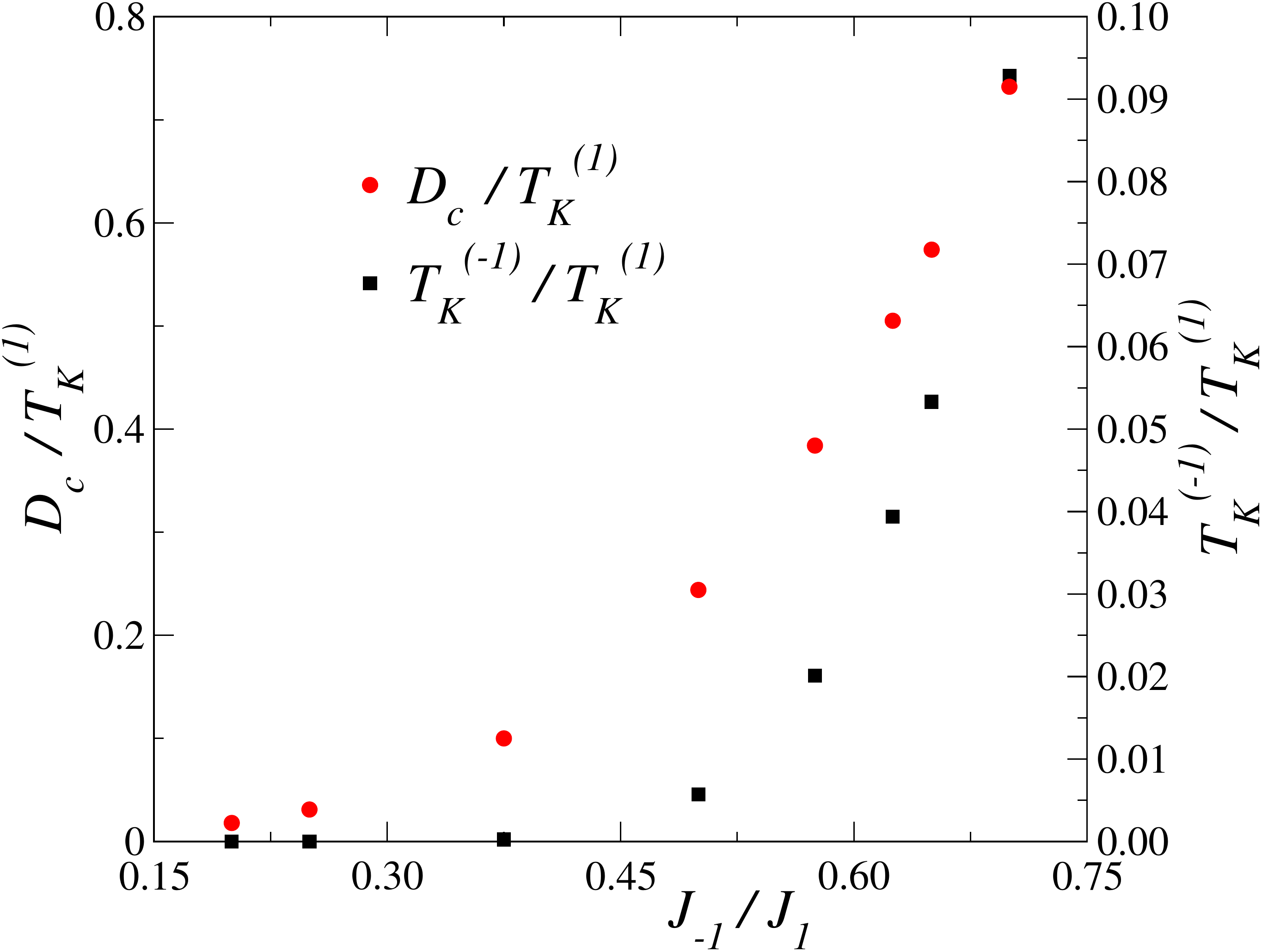}
\caption{
{\bf Characteristic energy scales}. Ratio of the lower Kondo temperature $T_K^{(-1)}$ over the higher Kondo temperature
$T_K^{(1)}$ (right scale,
red symbols) and ratio of the critical anisotropy $D_c$ over higher
Kondo temperature (left scale, black squares), as a function of the
smaller exchange coupling $J_{-1}$ at constant larger coupling $J_1=0.4$.}
\label{tkrd}
\end{center}
\end{figure}

We define the Kondo temperature $T_K^{(\tau)}$ as the temperature for which the contribution of
channel $\tau$ to the zero-bias conductance for $D=B=0$ falls to half of its zero-temperature value \cite{blesio18}.
For the Kondo model with channel symmetry
[$J_{-1}=J_1$ in Eq.~\eqref{hk}], the
topological transition takes place for $D_c \sim 2.5 T_K^{(1)}$ \cite{blesio18}, but $D_c$
decreases with decreasing $J_{-1}/J_1$ ratio. In fact, for $J_{-1}=0$ it is
known to be zero \cite{cornaglia11}.
In FePc experimentally $D =5$ meV \cite{hira} and $T_K^{(1)} \sim 20$ meV \cite{mina}.
In Fig. \ref{tkrd} we represent the ratios $D_c/T_K^{(1)}$ and $T_K^{(-1)}/T_K^{(1)}$
for intermediate values of $J_{-1}/J_1$. 
The higher Kondo temperature $T_K^{(1)}$ changes only moderately
as $J_{-1}/J_1$ varies, from 0.024 for $J_{-1}/J_1=0.2$ to 0.018 for $J_{-1}/J_1=0.7$
as a consequence of the competition between both channels \cite{joaq}. 
Instead, $D_c$ and particularly $T_K^{(-1)}$ have a strong dependence on $J_{-1}$
in the range considered.

For comparison with experiment we take $D/W=0.01$ (that means $W \sim$ 500 meV~\cite{gao07}, a different choice of $W$ does 
not affect the results in a sensitive manner) and fix $J_1=0.44$ so that 
$T_K^{(1)} \sim 200$ K as in experiment. 
The remaining parameter was fixed at 
$J_{-1}=0.22$ so that the system is in the non-Landau phase ($D>D_c$), 
with the width
of the dip near to the experimentally observed one (as described later).
The resulting Kondo temperatures are $T_K^{(1)}=198$ K and $T_K^{(-1)}=1.77$ K, 
while $D_c=0.00950 \sim 4.7$  meV $\sim 55$ K.

We choose as the corresponding parameters in the auxiliary Anderson model $J_H=0.1$,
$U_\tau=0.4$, $\epsilon_\tau=-U_\tau/2$, $\Delta_1=0.06$, $\Delta_{-1}=0.0345$.
The two models have the same low-energy behavior if the results are rescaled in terms of $T_K^{(1)}$ or $D_c$.

\subsection*{Differential conductance at zero temperature in the absence of magnetic field}

\begin{figure}[ht]
\begin{center}
\includegraphics*[width=\columnwidth]{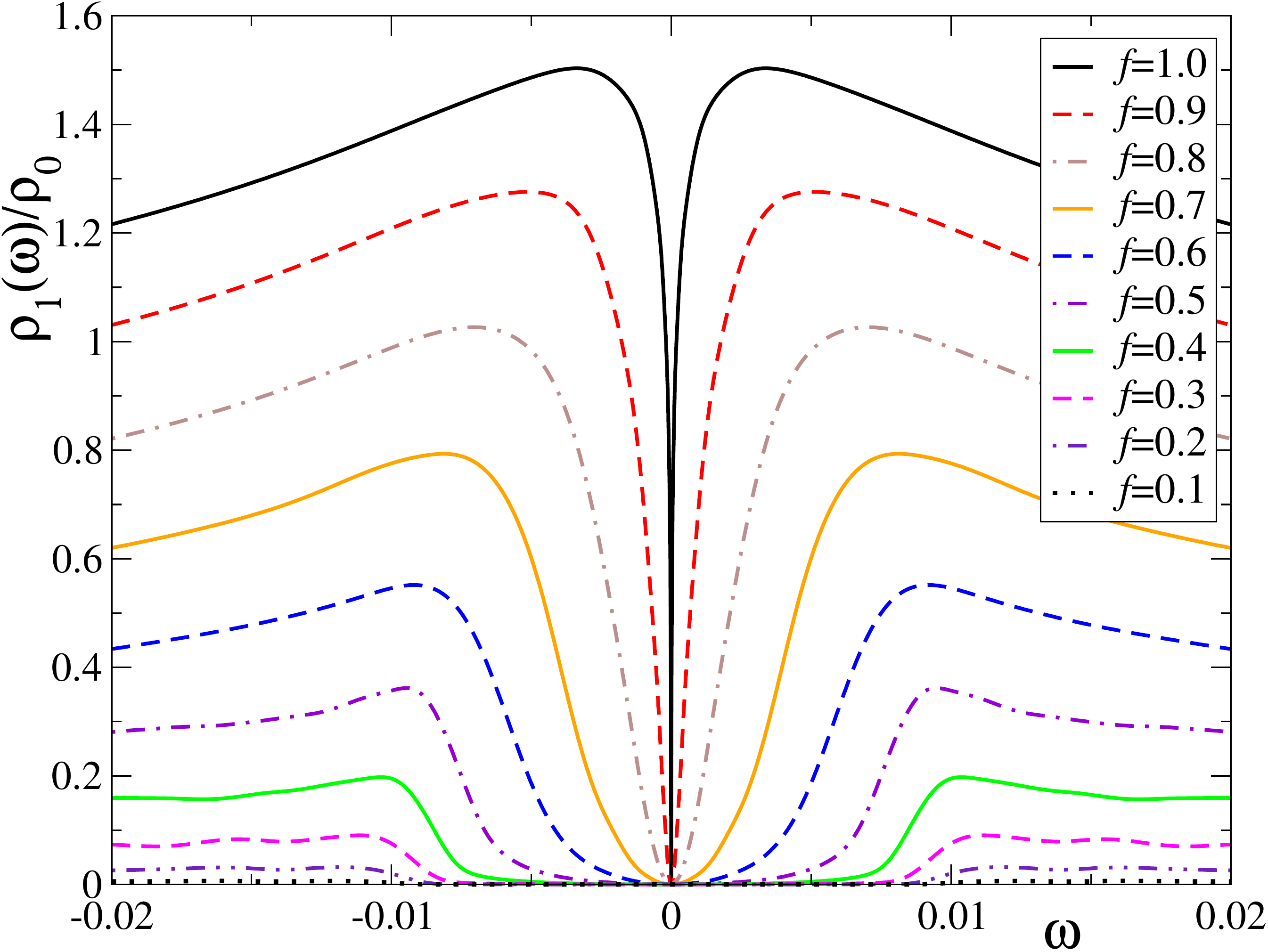}
\caption{{\bf Variation of the spectral density as the molecule is
raised from the surface.} The gradual decoupling
from the surface is modelled through equal reduction of both exchange coupling constants
by a factor $f$.
The model parameters are $J_1=0.44\times f$, $J_{-1}=0.22\times f$ and $D=0.01$.
See Supplementary Fig. 3} for a direct comparison with experiments.
\label{factor}
\end{center}
\end{figure}

The $s$-like orbitals of the STM tip have a larger overlap with
the $3z^2-r^2$ orbital of FePc than with the molecular $\pi$
orbitals and the substrate conduction-band wave functions. 
Ab initio calculations for Co on Cu(111) confirm this picture \cite{tacca}.
For this
reason, the dominant contribution to the experimental $dI/dV$ spectra corresponds to the spectral
density $\rho_1(\omega)=\rho_{1 \uparrow}(\omega)+\rho_{1 \downarrow}(\omega)$ in channel
$\tau=1$. This description is both simpler and more physically realistic than the assumptions
underlying the alternative interpretations of the measured $dI/dV$ from
Refs.~\onlinecite{mina,hira,yang}. The observed weak asymmetry in the line shapes indicates that some interference effects involving conduction orbitals are present. 
They will be incorporated in the next subsections. Here they do not modify the
essential features and the conduction orbitals become less important as the molecule
is raised from the surface.

Our result for FePc in the relaxed geometry
corresponds to the black full line of Fig.~\ref{factor} and reproduces the main features of the
observed differential conductance except for some asymmetry in the experiments, 
which we neglect in this subsection as explained above.
As the molecule is raised from the substrate by the attractive force of the STM tip, the Kondo
coupling to the substrate is reduced \cite{hira}. We assume that both exchange coupling constants
$J_\tau$ are reduced by the same factor $f$ 
because the same power-law dependence of the tunneling parameters $V_\tau$ with the distance is expected.
between the localized 3d orbitals and the extended conduction states \cite{harri}. As this
factor $f$ decreases from 1, the spectral density flattens and the dip broadens so that the
line-shape becomes similar to that observed in single-channel Kondo systems with magnetic
anisotropy \cite{zitko08,zitko10}. The exact same trend is observed experimentally
\cite{hira}. For very small $f$, there should be two sharp steps at the unrenormalized
threshold for inelastic spin-flip excitations ($\omega=\pm D$); in our
calculations they are overbroadened due to technical reasons that limit the resolution of the NRG
at large energies \cite{bulla}. A detailed comparison with experiment is contained in Supplementary Note 4.

It would be interesting to push the STM tip against the molecule and tune from the tunneling
to the contact regime. In this case the factor $f$ is expected to increase beyond 1
\cite{choi,lorente,diego}, driving the system through the topological quantum phase
transition to the ordinary Fermi liquid regime. This would be signalled by the sudden transformation 
of the dip into a peak of
similar
amplitude, i.e., compared to the baseline density of states, the
height of the peak just after the transition should be similar as the depth of the dip just before the transition.
This is an important prediction of our theory that should be the target of future experiments
(although in the contact regime, the widths of $dI/d(eV)$ and
$\rho_1(\omega)$ may differ due to non-equilibrium effects \cite{diego}). 

\subsection*{Temperature dependence}

\begin{figure}[ht]
\begin{center}
\includegraphics*[width=\columnwidth]{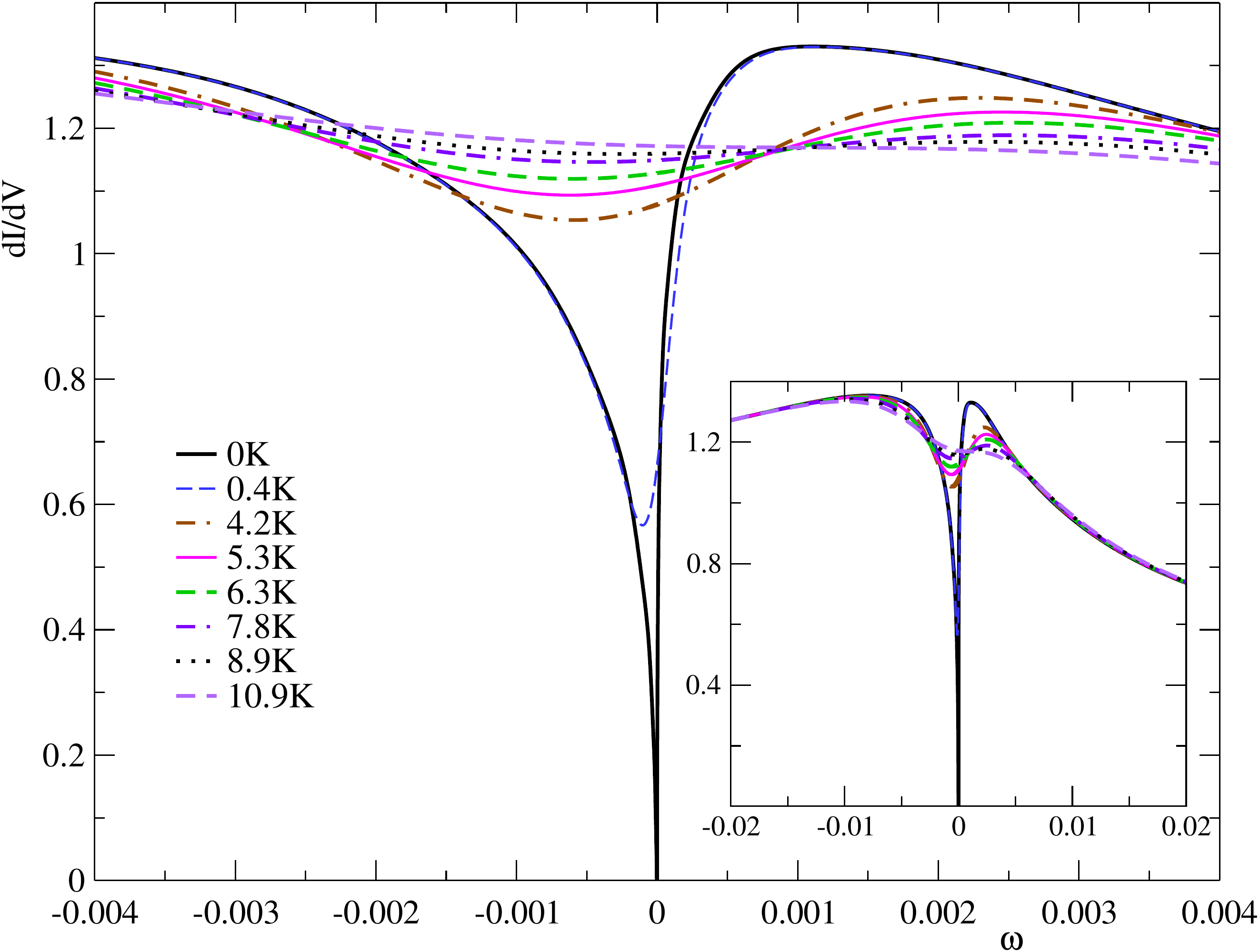}
\caption{
{\bf Temperature dependence of the differential conductance.} The main
panel shows the close-up on the low-energy regions, while the inset
shows an extended range.
Parameters as in Fig. \ref{factor}.
See Supplementary Fig. 4 for a direct comparison with experiments.}

\label{ft}
\end{center} 
\end{figure}

Assuming that the molecule is at equilibrium with the substrate, the
observed differential conductance is proportional to

\begin{equation}
\frac{dI(V)}{dV}\propto \int d\omega \rho _{h}(\omega )f^{\prime}(\omega
-eV),  \label{g}
\end{equation}
where $f^{\prime}(\omega )$ is the derivative of the Fermi function and 
$\rho_{h}(\omega )$ is the density of a mixed state which contains the orbitals
that have a non-vanishing hopping to the STM tip with an amplitude
proportional to the hopping \cite{sd} In the present case we take only two
contributions: the $3d_{3z^{2}-r^{2}}$ and the conduction states of the same
symmetry. Assuming that the latter corresponds to a flat band in the absence
of hybridization with the localized states,  one has \cite{zitkoq} 

\begin{equation}
\rho _{h}(\omega )\propto \Sigma _{\sigma }
\left[ (1-q^{2})\text{Im}G_{1\sigma }^{d}(\omega )
+2q\text{Re}G_{1\sigma }^{d}(\omega )\right] ,
\label{rhoh}
\end{equation}
where $q$ is proportional to the amplitude between the STM tip and the
conduction electrons of $3z^{2}-r^{2}$ symmetry and is responsible of the observed
asymmetry in the line shape. We take $q=0.4$, a similar value as taken in Refs.\onlinecite{mina,tsukahara11}, although the results do not show a high
sensitivity to $q.$

The temperature dependence of the dip is shown in Fig. \ref{ft}. The half width
of the dip at $T=0.4$ K, taken at the average between the minimum $dI/dV$ and
the relative maximum near $eV=0.001$ is 3.1 K,
near the reported one \cite{hira} 2.7 K. With
increasing temperature, the low-energy dip in the spectral density decreases
and eventually disappears, as expected. The effect is much more pronounced 
at low temperatures. The same trend is observed in the
experiment \cite{yang}. The inset shows the results in a scale of energies
similar to that of the experiment (Fig. 2 of Ref.~\onlinecite{yang}).
See Supplementary Note 4.

\subsection*{Magnetic field dependence}

\begin{figure} [ht]
\begin{center}
\includegraphics*[width=\columnwidth]{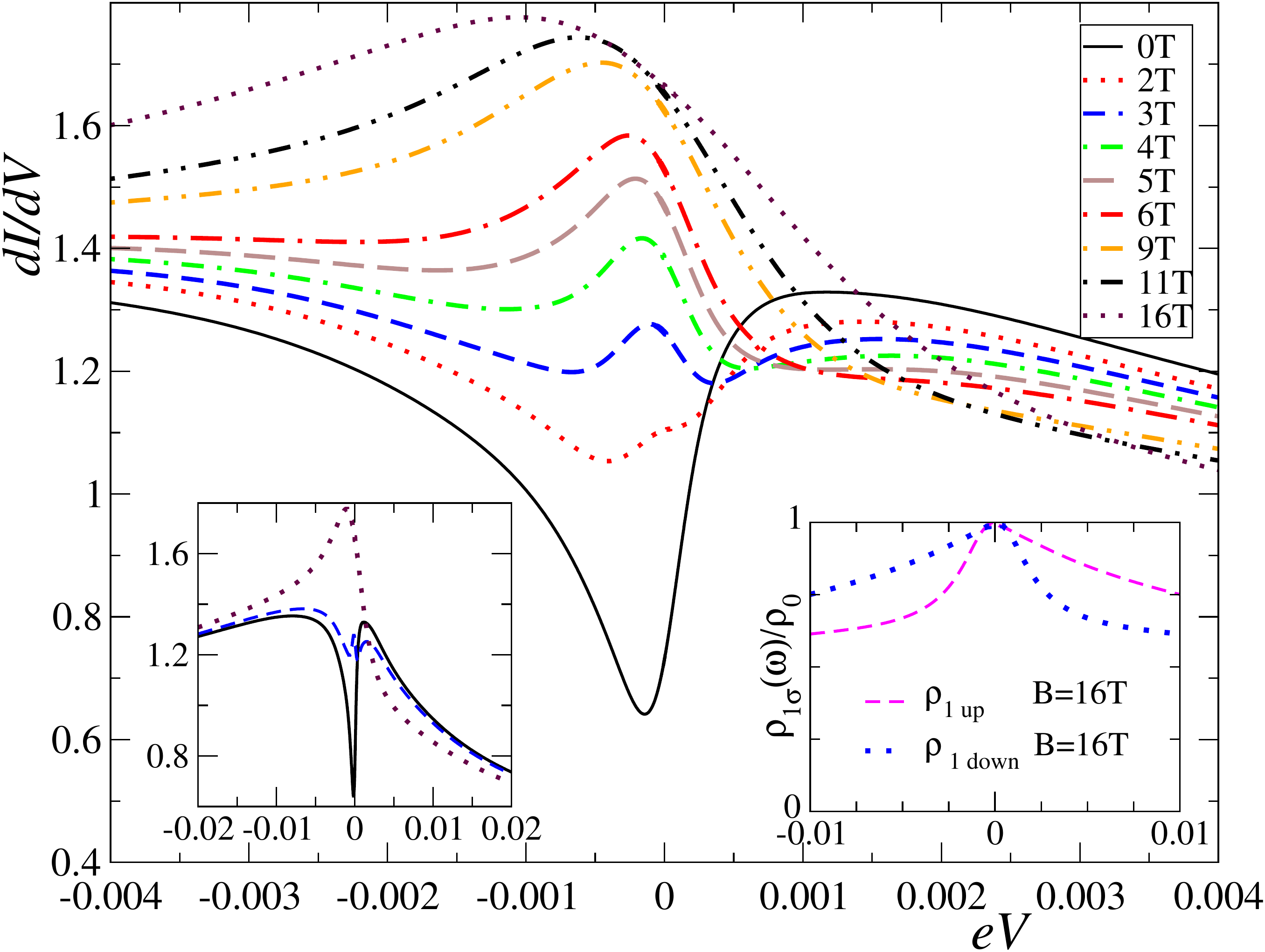}
\caption{{\bf Magnetic field dependence of the differential conductance.}
Left inset: differential conductance on a wider energy scale. 
Right inset: spin-resolved contributions to the spectral density of 
the $3d_{3z^{2}-r^{2}}$ states.
Parameters as in Fig. \ref{factor}, with the temperature $T=0.4$ K.
See Supplementary Fig. 5 for a direct comparison with experiments.
}

\label{fb}
\end{center}
\end{figure}

The effect of the magnetic field on the $dI/dV$ spectra reported in Ref.~\onlinecite{yang}
is most striking. A moderate field of $B=\unit[11]{T} \sim \unit[14]{K}$ readily transforms the
dip into a peak with a transition near 3 to 4 T. This phenomenon is fully reproduced in our calculations, see Fig.~\ref{fb}. For larger fields 
$B> 20$ T ($4.65 \times 10^{-3}$) a dip s expected to form
at the Fermi level as a consequence of the Zeeman splitting 
between $\rho_{1 \uparrow}(\omega)$ and $\rho_{1 \downarrow}(\omega)$.
For the comparison with experimment we have taken the giromagnetic factor 
$g=2$, inside the range of uncertainty of previously reported values (see 
supplemental material of Refs. \cite{hira,yang}).
A detailed comparison with experiment is contained in Supplementary Note 4.

The interpretation of this spectral transformation is highly non trivial. A field-induced
topological transition from a NLFL to an OFL would result in an abrupt change from a dip
into a peak, clearly at odds with the continuous evolution shown in Fig. \ref{fb} and observed in
the experiment. To shed more light on this discrepancy, we investigated the Luttinger integrals
entering Eqs.~\eqref{del}. This study was performed in terms of the equivalent auxiliary Anderson
model (see Methods) and interpreted using Eqs.~\eqref{tn}.

\begin{figure} [ht]
\begin{center}
\includegraphics*[width=0.9\columnwidth]{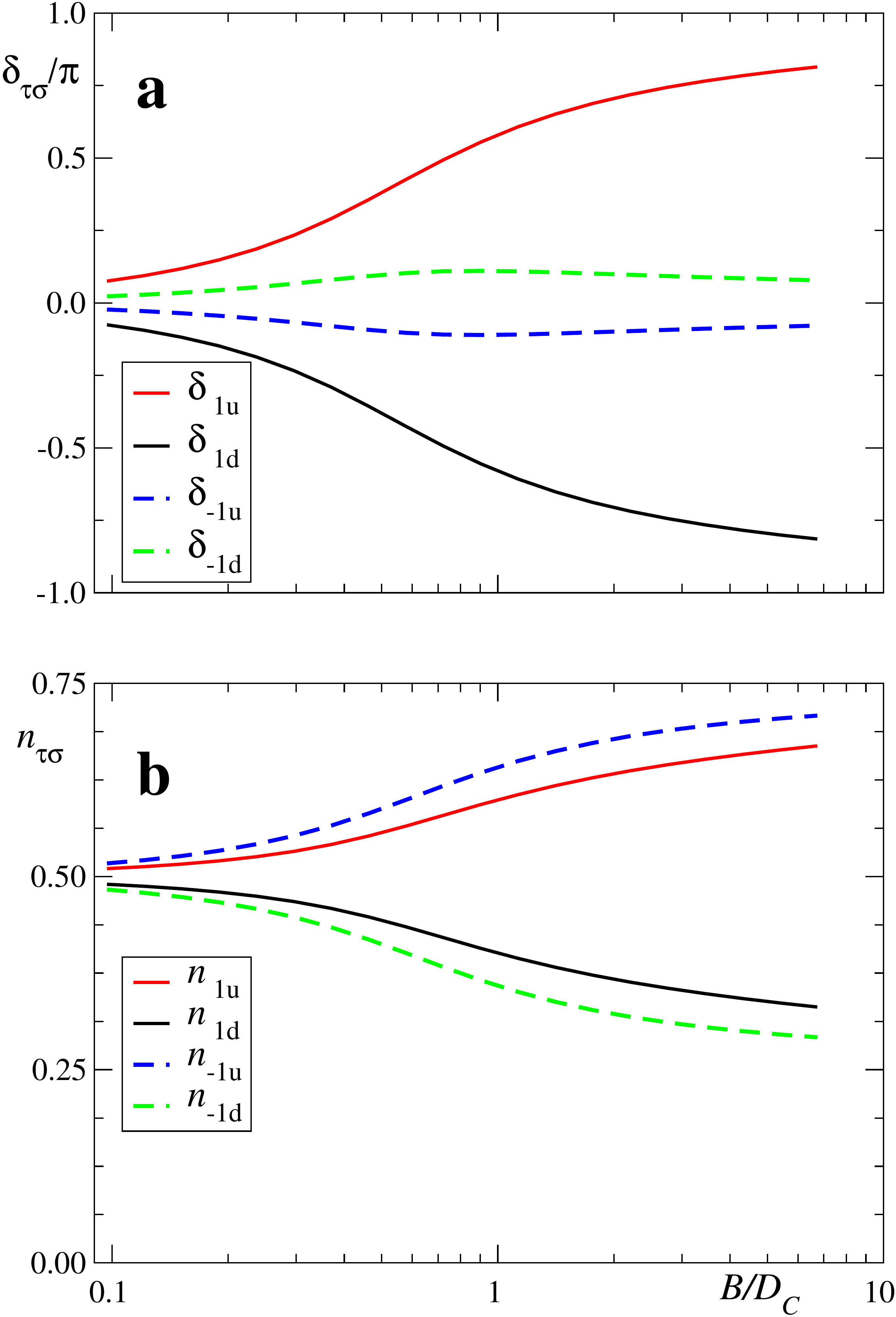}
\caption{{\bf Evolution of impurity scattering and occupancy
parameters.}
\textbf{a} Phase shifts $\delta_{\tau\sigma}$ and \textbf{b}
occupancies $n_{\tau\sigma}$ in the auxiliary 
Anderson model as a function of magnetic field $B$ for $D/D_c=1.67$.
$D_c$ is the anisotropy at the transition point.}
\label{delta}
\end{center}
\end{figure}

In Fig. \ref{delta} \textbf{a} we show the evolution of the phase shifts 
$\delta_{\tau \sigma }$ with
increasing field for $D/D_c>1$ (as in Figs.~\ref{factor},\ref{ft}, and \ref{fb}). 
For $B=0$, the four phase shifts 
$\delta_{\tau \sigma }=0$ (mod. $\pi$) and, therefore, the corresponding four Fermi-level 
spectral densities $\rho_{\tau \sigma }(0)=0$ [see Eqs. (\ref{rho})].
As $B$ increases, the phase shifts for the $\pi$ molecular orbital ($\tau=-1$) change only
moderately (below $0.12 \pi$) and therefore, the corresponding spectral densities
continue to exhibit a dip at the Fermi level. By contrast, the phase shifts that correspond to the
$3z^2-r^2$ molecular orbital ($\tau=1$) change considerably, reaching the values $\pm \pi/2$ for
$B/D_c \sim 0.8$. At this point the spectral densities 
$\rho_{1 \sigma }(0)$
reach their maximum value. Since $\rho_{1 \sigma }$ are the main contribution 
to the differential conductance, the above mechanism explains
the transformation from a dip into a peak
displayed in Fig.~\ref{fb}. Further increase of $B$ would lead to a Zeeman splitting,
as expected for $B \gtrsim T_K^{(1)}$: the peak in $\rho_{1 \uparrow}(\omega)$ displaces to lower
energies and that in $\rho_{1 \downarrow}(\omega)$ to higher energies, so that 
$\rho_{1 \sigma}(0)$ decrease for both $\sigma$, which corresponds to $|\delta _{1 \sigma }|$ increasing
beyond $\pi/2$ in agreement with Eq.~\eqref{rho}.

The changes in the spectral densities at the Fermi level with magnetic field should lead to important spatial variations of the differential conductance at small bias. 
For $B=0$ in the Kondo limit (integer total occupancy),
the spectral densities of the localized orbitals with $\pi$ and $3z^2-r^2$ 
symmetry vanish at the Fermi
surface and the space variation is dominated by conduction states. As $B$ increases, the influence of
$\pi$ orbitals remains small but that of the cylindrical symmetric $3z^2-r^2$ orbitals increase considerably and 
therefore, an evolution to a more circular shape is expected, as observed experimentally \cite{yang}.

\begin{figure}
\begin{center}
\includegraphics*[width=\columnwidth]{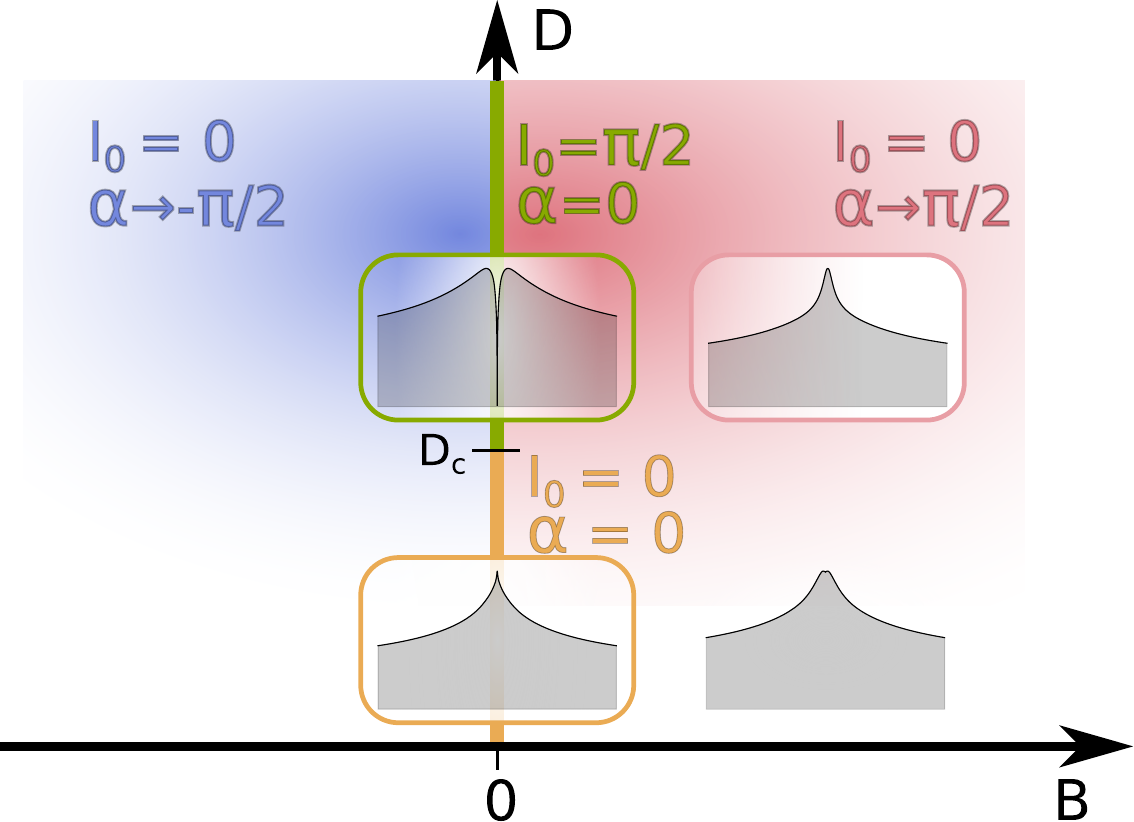}
\caption{{\bf Schematic phase diagram of the topological
properties.}
Phase diagram of the auxiliary Anderson model as a function of the magnetic field $B$ and the magnetic
anisotropy $D$.
Along the green half line $B=0$, $D>D_c$, the system 
is a non-Landau Fermi liquid (NLFL) with the topological value $I_0=\pi/2$, and the non-topological
parameter $\alpha=0$. For an infinitesimal $B$ both 
quantities jump as indicated in the figure, yet all physical quantities
are continuous. For $D<D_c$, $I_0=0$, and $\alpha$ is continuous 
across $B=0$, where its value crosses zero (orange segment). 
The insets show the low-energy region around the Fermi level:
regular Fermi liquid (orange frame) and NLFL (green frame) at
zero magnetic field, as well as weakly spin-polarized Fermi
liquids at moderate magnetic field (white and pink frame).
}
\label{phd}
\end{center}
\end{figure}

The topological properties under an applied magnetic field are very peculiar. For $B=0$ and $D >
D_c$, there is a pole in the
self-energy $\Sigma(\omega)$ on the real
axis at $\omega=0$.
An infinitesimal $B$
displaces this pole 
away from $\omega=0$, as it is shown in Supplementary Note 3. 
$I_0$ drops to zero and the system becomes an
OFL. However at the same time $\alpha$ jumps to $\pi/2$ ($-\pi/2$) for infinitesimal
positive (negative) $B$.  
See Fig. \ref{phd}.
As a result, in both cases the Luttinger integral $I_{-1 \sigma }$ for
the majority spin [that is $\uparrow$ ($\downarrow$) for positive (negative) $B$] remains $\pi/2$, while
$I_{1 \sigma }$ jumps from $\pi/2$ to $-\pi/2$, see Eqs.~\eqref{il2}. This jump in $\pi$ does not
affect any physical properties which are continuous across $B=0$, as expected and in-line
with the ``modulo $\pi$ ambiguity'' in the definition of the scattering phase shifts \cite{taylor}.
The Luttinger integrals for minority spin are obtained by interchanging the orbital index. 

\begin{figure}
\begin{center}
\includegraphics*[width=\columnwidth]{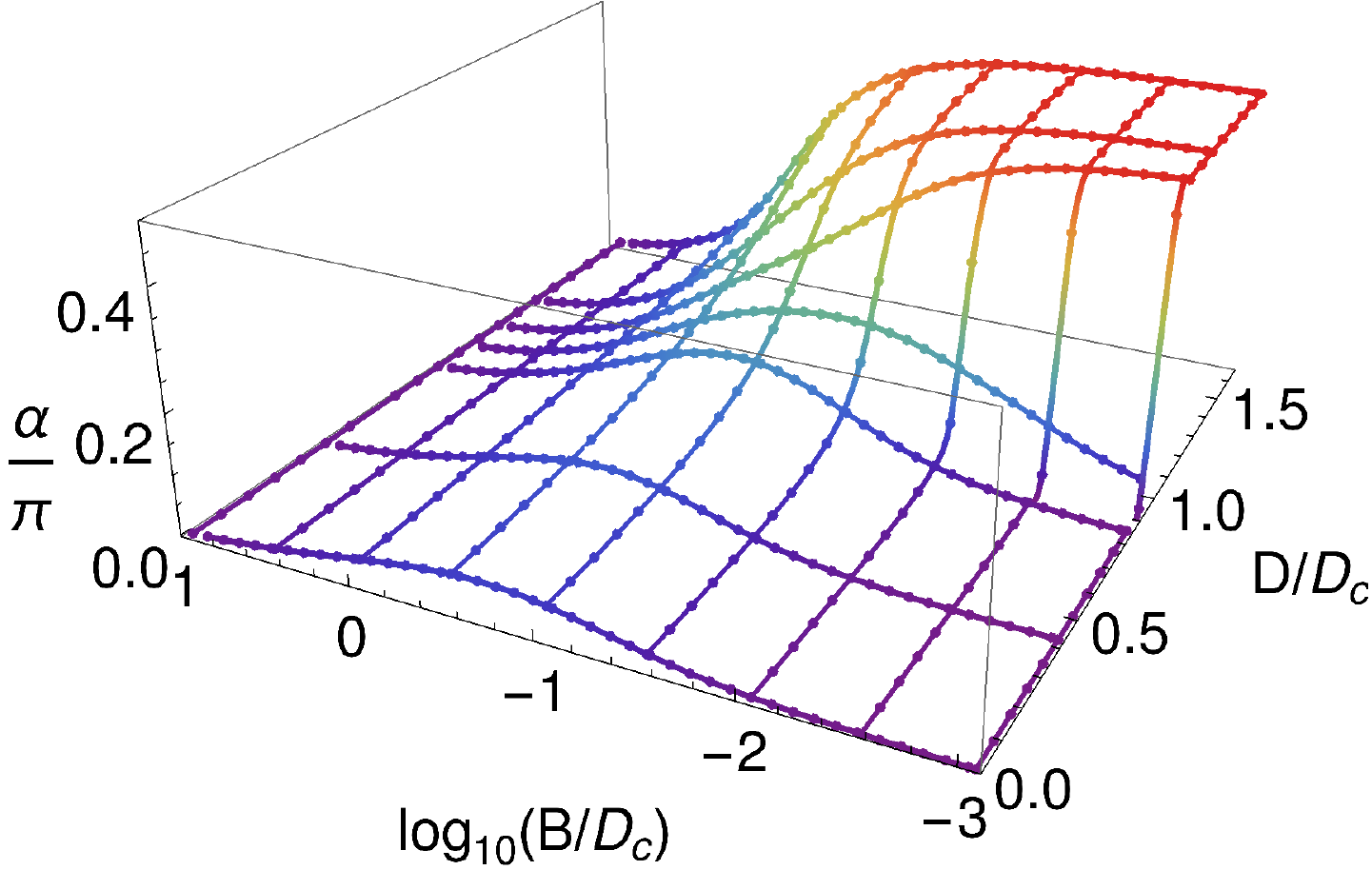}
\caption{{\bf Non-topological parameter.} Phase shift $\alpha(D,B)$ for the
auxiliary Anderson model as a function of magnetic anisotropy $D$
and magnetic field $B$. The color encodes the value of $\alpha$.}
\label{xd}
\end{center}
\end{figure}

For $D<D_c$, $I_0$ and $\alpha$ are continuous as functions of $B$, and both are equal to zero
for $B=0$. In fact, in the
whole $(D,B)$ plane, $I_0 \equiv 0$ and $\alpha$ is continuous except on the half-line $B=0$, $D>D_c$.
This line is, hence, a branch-cut for $\alpha$. 
See Fig. \ref{phd}.
Furthermore, the point $D=D_c$, $B=0$ may be
considered as a logarithmic singularity for the function $\alpha(D,B)$ viewed as a function in the complex
plane with the argument $z=D+iB$.
In Fig. \ref{xd} we display $\alpha$ as a function of both 
$D/D_c$ and  $B/D_c$.
As explained above for $B\rightarrow 0^+$,
$\alpha(D,B=0)=(\pi/2)\theta(D-D_c)$, where $\theta(x)$ is the step function. For small non-zero $B$,
$\alpha(D,B)$ still resembles this step function. In fact, $\alpha(D,B) \sim
-\frac{1}{2}\mathrm{Im}\ln[D_c-D-iB]$.
As $|B|$ further increases, $\alpha(D)$ decreases markedly for $D > D_c$.
For $|B| \gg D_c$, $\alpha$ is small for all $D$ and as a consequence also the four Luttinger integrals
are small.

As $|B|$ increases, the occupancies $\left\langle n_{\tau \uparrow }\right\rangle$ increase
[see Fig. \ref{delta} \textbf{b}],
with the effect of increasing $\delta_{\tau \uparrow }$ [see Eqs. (\ref{del})].  
For $\delta_{-1
\uparrow }$, this effect is largely compensated by the decrease in $\alpha$, leading to a small
overall variation of the phase shift. Instead, for $\delta_{1 \uparrow }$ both effects
add up and $\delta_{1 \uparrow }$ has a marked increase, which is then reflected in the
transformation of the dip into a peak. For spin $\downarrow$ the effects are opposite.

\section*{Discussion}

The ability to flip the differential conductance between low and high values by
tuning some external parameter is of tremendous practical importance in molecular electronics
for switching device applications. Recently a topological quantum phase transition has been found in 
a spin-1 Kondo model with two degenerate channels and single-ion magnetic anisotropy, 
in which the zero-bias differential
conductance and the spectral density at zero temperature jump from their maximum possible values
to zero as the longitudinal single-ion anisotropy $D$ is increased beyond a threshold
value $D_c$ \cite{blesio18,blesio19}. 
By generalizing the theory of such topological transitions to nonequivalent channels and finite magnetic 
field, we have for the first time conclusively identified FePc/Au(111) as an experimental realization of 
this phenomenon, since our approach provides a unified description of the totality of experimental 
observations. 
The main difficulty in this identification is that for
$D>D_c$ the system is an unconventional Fermi liquid that cannot be adiabatically
connected to a non-interacting system by turning off the interactions
\cite{blesio18,blesio19,curtin18,nishi}. The corresponding concept of a non-Landau Fermi
liquid (NLFL) remains largely unfamiliar to most of the physics community. In the NLFL, the
Friedel sum rule has to be generalized by introducing a topological quantity that has been
previously overseen, even though it leads to a dramatic drop in spectral density and
differential conductance for $D>D_c$.

By formulating a general theory of the topological aspects of the Friedel sum rule that
incorporates the effects of broken channel and spin symmetries, fully corroborated by numerical
calculations for a model Hamiltonian, we reliably established that FePc on Au(111) behaves as a
NLFL. With a single set of parameters we explained three key experiments on the differential
conductance of this system \cite{mina,hira,yang}. The sole assumption (i.e., that the scanning tunneling
microscope senses mainly the 3d orbital of Fe with $3d^2-r^2$ symmetry) is simple, physically realistic, and consistent with ab initio calculations \cite{tacca}

The same model also explains the recent experiments for MnPc on Au(111) \cite{guo}, 
which exhibit a dependence on magnetic field with qualitative features similar to those in FePc on Au(111).
Alternative explanations proposed for these experiments 
are questionable or contradict well established facts,
like the role of Hund rules in defining ground-state multiplets.

While a moderate magnetic field $B$ of the order of \unit[10]{T} leads to a continuous transition 
from very small to very high zero-bias conductance, we predict an abrupt transition for $B=0$
if the tip is pressed against the molecule changing the regime of the system from the tunneling 
to the contact one, thereby increasing the Kondo temperatures \cite{choi,lorente,diego}.

\section*{Methods}

The numerical calculations were performed with the NRG Ljubljana \cite{zitko09,nrglj} implementation of the numerical 
renormalization group method \cite{bulla,wilson} using
the separate conservation of the isospin (axial charge) in each channel $\tau$, 
as well as the conservation of the $z$-component of the total spin, i.e., $SU(2) \times SU(2) \times
U(1)$ symmetry. The calculations for the Kondo model were performed with the discretization parameter 
$\Lambda=4$ with the broadening parameter $\alpha=0.8$ for Figs. \ref{ft} and \ref{fb}.
For Fig. \ref{factor} the values of $\alpha$ used were 0.1 for $f \leq 0.1$, 
0.2 for $f=0.6$ and 0.7, 0.4 for $f=0.8$ and 0.9 and 0.6 for $f=1$.
The calculations for the Anderson model have been performed using 
$\Lambda=3$ with the broadening parameter $\alpha=0.3$.
We kept up to 10000 multiplets (or up to cutoff 10 in energy units) in the truncation, 
averaging over $N_z=4$ different discretization meshes. The spectral functions were computed using 
the complete Fock space algorithm \cite{peters06}, and the resolution for the Anderson model 
was improved using the ``self-energy trick'' \cite{bullaself}. 

For large interaction $U$,  the value of $D_c$ calculated with the 
Anderson model Eq. (\ref{ha}) coincides with that obtained from the 
corresponding Kondo model,
However, for example for $U=4$, we obtain that the phase shift modulo $\pi$ calculated directly from the NRG spectrum and that obtained the using the 
generalized Friedel sum rule Eqs. (\ref{del}) deviate by up to $0.1 \pi$ for large $B$. 
This is due to the fact that Eqs. (\ref{del}) were simplified for the case in which
the conduction band width (which we have taken as $2W=2$) is much larger than all other energies
involved (the ``wide-band limit''), in such a way that the number of conduction electrons per channel is not modified by the
addition of the impurity \cite{lang}. To avoid this difficulty, we have used 
$U=0.4$ in the calculations with the Anderson model. In this case the maximum 
deviation between both phase shifts is below
$0.02 \pi$ and the critical anisotropy is reduced 
by a factor of the order of 6.
However, the physical properties are very similar for the same $D/D_c$ and $B/D_c$.

\section*{Data availability}

The datasets generated during the current study are available in the
Zenodo repository under accession code \url{10.5281/zenodo.5506654}. The
data includes figure sources and model
definition files for the NRG solver.

\section*{Code availability}

The NRG calculations presented in this work have been performed with
the NRG Ljubljana code. The source code is available from GitHub,
\url{https://github.com/rokzitko/nrgljubljana}. A snapshot of the
specific
version used (8f90ac4) has been posted on Zenodo,
\url{http://doi.org/10.5281/zenodo.4841076}.

\section*{Acknowledgments}

R\v{Z} acknowledges the support of the Slovenian Research Agency (ARRS) under P1-0044.
GGB and LOM are supported by PIP2015 No. 364 of CONICET
AAA is supported by PIP 112-201501-00506 of CONICET, 
and PICT 2017-2726 and PICT 2018-01546 of the ANPCyT, Argentina.

\section*{Author contributions}
A. A. A. and L. O. M. conceived the project, R. \v{Z}. and G. G. B. performed the NRG calculations. 
A. A. A., R. \v{Z}. and L. O. M. wrote the paper.

\section*{Competing Interests}
The authors declare no competing interests.

\clearpage

\appendix

\setcounter{figure}{0}
\renewcommand\thefigure{S\arabic{figure}}


\begin{widetext}

\section*{Supplementary Note 1: Auxiliary multiorbital two-channel Anderson model}
 
In order to analyze the topological features of the $S=1$ Kondo model (1) with two non-equivalent channels, encoded in
the generalized Friedel 
sum rules that we will deduce in Supplementary Note 2,  it is better to work with the auxiliary multiorbital two-channel Anderson Hamiltonian, 
\begin{equation}
H_A = H_0 + H_{\rm int}. 
\label{hamilton}
\end{equation}
$H_0$ contains the one-body terms:
\begin{equation}
H_0 = H_{\rm cond} + H_{\rm hyb} + H_{d}.   
\end{equation}
$H_{\rm cond}$ corresponds to two conduction bands (channels): one with symmetry $3z^2-r^2$ ($\tau=1$) and another that takes into account the degenerate $\pi$-manifold $xz,yz$ ($\tau=-1$)~\cite{hira}:
\begin{equation}
H_{\rm cond}  =  \sum_{k \tau \sigma}\varepsilon_{k \tau} {c}^\dagger_{k \tau \sigma}{c}_{k \tau\sigma}; 
\end{equation} 
we consider conduction electron energy dispersions such that both bands have the same constant density of states (DOS) and
we work in the wide-band limit, such that the half-bandwidth $W$ is the largest energy scale in the problem.

$H_{\rm hyb}$ contains the hybridization between the conduction and impurity states with the same $\tau$ symmetry: 
\begin{equation}
H_{\rm hyb}   =   \sum_{k \tau}\left(V_{ \tau}{c}^\dagger_{k \tau \sigma}
{d}_{\tau \sigma}+ {\rm H. c.}\right).   
\end{equation}
For correct description of FePc on Au(111), it is essential to allow for very different coupling strenghts $V_\tau$ 
of each localized orbital with the substrate, giving rise to a multiorbital  Anderson model with non-equivalent orbitals.
Due to its spatial dependence, the $\tau=1$ localized orbital hybridizes stronger with the gold surface than the
$\tau=-1$ orbital. As a consequence, $V_{1} > V_{-1}$.

$H_{d}$ includes the impurity single-orbital (hole) energies and the Zeeman term:
\begin{equation}
H_{d}  = \sum_{\tau\sigma}\epsilon_{\tau}  n_{\tau\sigma} - B S_z \equiv 
\sum_{\tau\sigma}\epsilon_{\tau\sigma} n_{\tau\sigma},\;\;{\rm with}\;\;\epsilon_{\tau\sigma} \equiv \epsilon_{\tau}- \sigma\frac{B}{2}.
\end{equation}
We consider a Zeeman term only at the impurity site because for constant conduction-band DOS in
the wide-band limit, the magnetic field has no effect on the conduction band (i.e., it can be eliminated through a
simple energy shift).

The interaction terms act on the impurity states:
\begin{equation}
 H_{\rm int} = \sum_{\tau} U_\tau n_{\tau\uparrow}n_{\tau \downarrow} - J_H{\bm S}_{1}\cdot {\bm S}_{-1}+ D \left(\sum_{\tau} {S}_{z \tau}\right)^2.
 \label{hint}
\end{equation}

We study this model in the particle-hole symmetric case, corresponding to $\epsilon_{\tau} = - \frac{U_\tau}{2}$ and
symmetric conduction bands relative to the $\omega=0$ Fermi level) in order to have exactly one hole in each impurity orbital. In the Kondo regime, $-\epsilon_{\tau} \gg V^2_\tau/W$, these 
two holes are ferromagnetically coupled by the first Hund's rule, $J_H > 0$, yielding an effective $S=1$ at the impurity site. 
Due to symmetry considerations of the involved 3d orbitals, we take the same Hubbard repulsion $U$ for each impurity orbital. 
This implies that both single-orbital energies $\epsilon_\tau$ should be degenerate in the particle-hole symmetric case, while {\it ab initio} calculations~\cite{mina} give  
an energy difference of the order of 1 eV. In the Kondo regime this difference is irrelevant. 

Among the interactions of a generic multiorbital Anderson Hamiltonian for 3d impurities~\cite{oles83}, 
we are neglecting the inter-orbital Hubbard repulsion $U' n_{1} n_{-1}$, as this term is almost constant in the (particle-hole symmetric) 
Kondo regime. We  also neglect the pair hopping interaction $J_{H} ({d}^\dagger_{1 \uparrow}{d}^\dagger_{1 \downarrow}{d}_{-1 \downarrow}{d}_{-1\uparrow} + {\rm H.c.})$, 
as, for FePc on Au(111), the two-electron low energy sector contains only the triplet  states due to the strong Hund's coupling~\cite{mina}. 
While the inter-orbital Hubbard repulsion $U'$ could be easily 
included in the NRG computation, the pair hopping interaction breaks the separate conservation of the isospin in each channel $\tau$, resulting in a 
very demanding computational effort.

By means of the equation of motion method, it can be shown~\cite{hewson97} that all the conduction band information needed for the computation of impurity observables is encoded in the 
hybridization functions 
\begin{equation}
 \Delta_\tau (\omega) = \pi \sum_k V_{\tau}^2 \delta(\omega-\varepsilon_{k\tau}).
\end{equation}
As we assume equal constant conduction-band DOS for boths channels and energy-independent $V_\tau$, the hybridisation
functions $\Delta_\tau$ are likewise constant functions.
This is a very reliable assumption for normal conduction bands with a finite DOS at the Fermi level, as in this case the
Kondo physics depend only marginally on the exact form 
of the DOS. It should be stressed that, if we were interested in substrate observables like STM spectra far from the impurity, there would be a need for a more 
thorough modeling of the conduction bands and the tunneling parameters $V_\tau$.

It is worth to mention that, in the Kondo regime, the Anderson Hamiltonian (\ref{hamilton}) can be exactly mapped to
the Kondo Hamiltonian (1) by means of the Schrieffer-Wolff transformation~\cite{hewson97}, with Kondo exchange
interactions given by $J_\tau = 4V_\tau^2/U$. Potential scattering is absent due to the particle-hole symmetry. We have
numerically verified the agreement between the predictions for both Hamiltonians. Although the Kondo model is simpler
because of its reduced local Hilbert space, it is better to work with the auxiliar Hamiltonian (\ref{hamilton}) in
order to clearly interpret the Friedel sum rules, which involves the computation of the localized spin-orbital
occupancies $\langle n_{\tau \sigma} \rangle$.

\subsection*{Conservation laws for the Anderson Hamiltonian}

The Hamiltonian $H_A$ commutes with the total electron number {\it per} orbital
\begin{equation}
 N_{\tau} = \sum_{k \sigma}{n}_{k \tau \sigma} + \sum_{\sigma}{n}_{\tau\sigma} 
\end{equation}
($n_{k\tau\sigma} \equiv c^\dagger_{k\tau \sigma}c_{k\tau\sigma}$)
 and with the total electron number {\it  per} spin 
 \begin{equation}
 {N}_{\sigma} = \sum_{k \tau}{n}_{k \tau \sigma} + \sum_{\tau}{n}_{\tau\sigma}.  
\end{equation}
Consequently, the system conserves the total electron number, 
\begin{equation}
 N = \sum_{\tau\sigma}{N}_{\tau\sigma}, 
\end{equation}
the total spin projection along the $z$ direction, 
\begin{equation}
 {S}^{\rm tot}_{z} = \frac{1}{2}\sum_{\tau\sigma}\sigma {N}_{\tau\sigma}, 
 \label{szcons}
\end{equation}
and the total orbital isospin
\begin{equation}
 {T}_z = \sum_{\tau \sigma} \tau N_{\tau\sigma}.   
 \label{tzcons}
\end{equation}

As there are no spin-orbit couplings and the magnetic field $B$ is along the single-ion anisotropy axis $z$,  the Hamiltonian $H_A$ also 
commutes with the total spin 
\begin{equation}
{\bm S}^2 = \frac{1}{2}\left({S}_+ {S}_-+ {S}_- {S}_+ \right) + \left({S}_z \right)^2,   
\end{equation}
where 
$S_+ = \sum_\tau \left(\sum_{k} c^\dagger_{k\tau \uparrow}c_{k \tau \downarrow} + d^\dagger_{\tau \uparrow}d_{\tau \downarrow}\right), $
$S_- = (S_+)^\dagger$.
It is worth to mention that if the single-particle energies $\varepsilon_{k \tau \sigma}$ and $\epsilon_{\tau\sigma}$ in $H_0$ were arbitrary,
the total spin ${\bm S}^2$ would commute with $H_A$ only if the relations  
$\varepsilon_{\mu \tau \uparrow} - \varepsilon_{\mu \tau \downarrow} = \varepsilon_{\mu' \tau' \uparrow} - \varepsilon_{\mu'  \tau' \downarrow} $
hold for all $\mu, \mu' = k, d; \tau,\tau'= 1,-1$.  

\section*{Supplementary Note 2: Generalized Friedel sum rules} 

\subsection*{Green's functions and electron occupation numbers}
Under the conservation laws (\ref{szcons}), (\ref{tzcons}), it can be shown by means of the Lehmann representation that all the single-particle Green's functions (GF) of  $H_A$ , 
$$G^{k k'}_{\tau \sigma \tau'\sigma'} \equiv <\!\!\!<c_{k\tau\sigma}|c^\dagger_{k'\tau'\sigma'}>\!\!\!>,\;\; G^{k  d}_{\tau \sigma \tau'\sigma'} \equiv  <\!\!\!<c_{k\tau\sigma}|d^\dagger_{\tau'\sigma'}>\!\!\!>, \;\;$$
$$G^{d k'}_{\tau \sigma \tau'\sigma'} \equiv <\!\!\!<d_{\tau\sigma}|c^\dagger_{k'\tau'\sigma'}>\!\!\!>,\;\; 
G^{d}_{\tau \sigma \tau'\sigma'} \equiv <\!\!\!<d_{\tau\sigma}|d^\dagger_{\tau'\sigma'}>\!\!\!>,$$  are diagonal in spin and orbital indices. Thus, the GF matrix is block diagonal: 
\begin{equation}
 \bm{\mathbb G} = \left(\begin{array}{cccc} \bm{\mathbb G}_{1\uparrow} &   0 & 0 & 0 \\ 0 &  \bm{\mathbb G}_{1\downarrow} & 0 & 0 \\ 
 0 & 0 &  \bm{\mathbb G}_{-1\uparrow} & 0 \\ 0 & 0 & 0  &  \bm{\mathbb G}_{-1\downarrow}\end{array}\right), 
 \label{gmatrix}
\end{equation}
where the block indices run over the conduction electron momenta and the impurity, that is, over the set $I \equiv \left\{k_1, k_2, \cdots , d\right\}$.

As the interaction terms involve only the impurity degrees of freedom, we can express the conduction-band and mixed
Green's functions in terms of the impurity GF $G^d$ using the equations of motion.
We have \begin{eqnarray}
 G^{k k'}_{\tau  \sigma} & = & g_{k \tau} + |V_\tau|^2 g_{k \tau} G^d_{\tau\sigma} g_{k'\tau}, 
 \nonumber
 \\
 G^{k d}_{\tau \sigma} & = & V_{\tau} g_{k\tau} G^d_{\tau\sigma}, \label{eom}\\
 G^{dk'}_{\tau \sigma} & = & V^*_{\tau} g_{k' \tau} G^d_{\tau\sigma},
 \nonumber
\end{eqnarray}
where 
\begin{equation}
 g_{k \tau}  = \frac{1}{z-\varepsilon_{k \tau}}.
\end{equation}
The total electron numbers {\it per} spin and {\it per} orbital can be expressed as the complex-plane contour integrals~\cite{mahan}
\begin{equation}
 \langle {N}_{\tau\sigma}\rangle = \oint_{\Gamma} \frac{dz}{2\pi i} n_F(z) {\rm Tr} \; {\bm {\mathbb G}}_{\tau\sigma}(z),  
\end{equation}
where $n_F$ is the Fermi function and the contour $\Gamma$, shown in
Supplementary Fig.~\ref{contours},
encloses all the poles of the GF matrix (along the real axis) and none of the poles of $n_F$ (along the imaginary axis). 
\begin{figure} [ht] 
\begin{center}
\includegraphics*[width=0.49\columnwidth]{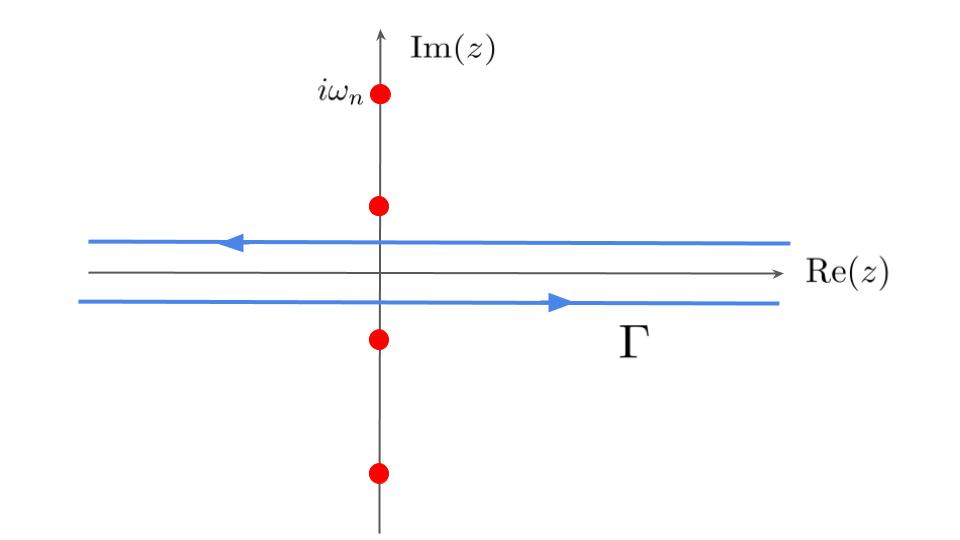} 
\hspace*{0.05cm} 
\includegraphics*[width=0.49\columnwidth]{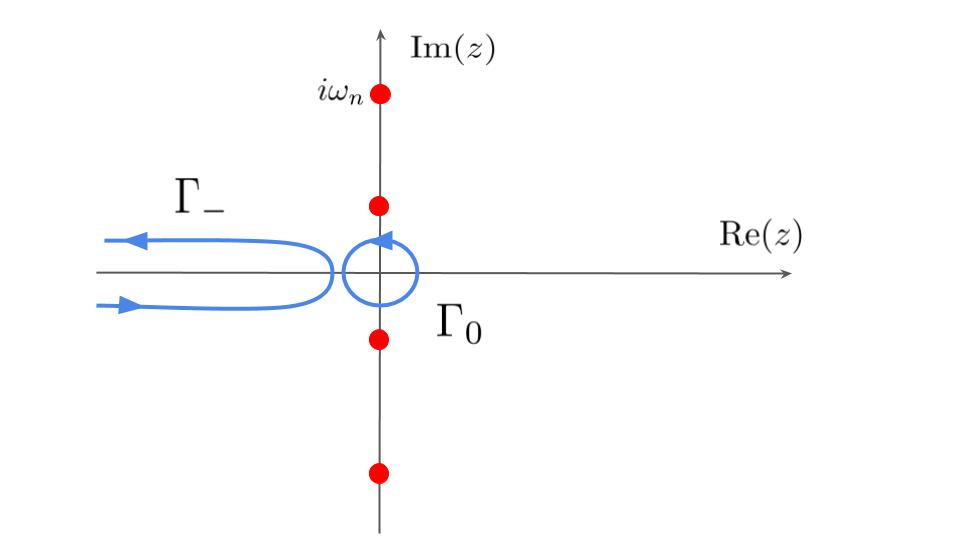} \\
\caption{{\bf Integration contours in the complex plane.} \textbf {a)}  $\Gamma$ encloses all the real axis; \textbf{b)} $\Gamma_0$, the origin of the
complex plane, and $\Gamma_{-}$, the negative real axis. The dots represent the fermionic Matsubara frequencies $i \omega_n =  i\frac{(2n+1)\pi}{\beta}$.}
\label{contours}
\end{center}
\end{figure}
Using (\ref{gmatrix}) and (\ref{eom}), we get 
\begin{equation}
 {\rm Tr}\;{\bm {\mathbb G}}_{\tau\sigma}(z) =  \sum_{k} G^{kk}_{\tau\sigma}(z) + G^d_{\tau\sigma}(z) = 
 \sum_{k}g_{k \tau} + \sum_{k} \frac{\left|V_{\tau}\right|^2}{(z-\varepsilon_{k \tau})^2} G^d_{\tau\sigma}(z) + 
 G^d_{\tau\sigma}(z). 
\end{equation}
If we define the complex hybridization function 
\begin{equation}
 \Gamma_{\tau}(z) \equiv \sum_{k}\frac{\left|V_{\tau}\right|^2}{z-\varepsilon_{k \tau}},
\end{equation}
then 
\begin{equation}
 {\rm Tr}\;{\bm {\mathbb G}}_{\tau\sigma}(z) =  \sum_{k}\frac{1}{z-\varepsilon_{k \tau}} + 
 \left(1-\frac{\partial \Gamma_{\tau}(z)}{\partial z}\right)G^d_{\tau\sigma}(z).
\end{equation}

Now we re-express the trace of the GF matrix in terms only of impurity functions (GF and self-energy).  
Using the formula~\footnote{By the method of equation of motion we obtain the relation 
$$
 G^{k k'}_{\tau \sigma} -  G^{kd}_{\tau \sigma}\frac{1}{G^d_{\tau \sigma}}G^{dk'}_{\tau \sigma} = g^k_\tau\delta_{kk'}.
$$
Therefore, the diagonal matrix with elements $g^k_\tau \delta_{k k'}$ is equal to  ${\mathbb G}_{\tau \sigma}/G^d_{\tau\sigma}$, that is, the Schur complement of 
$G^{d}_{\tau \sigma}$~\cite{horn12}. Supplementary Eq. (\ref{determinante}) follows straightforwardly from 
the Schur decomposition property $\det \mathbb{G}_{\tau \sigma} = \det \left(\mathbb{G}_{\tau\sigma}/G^d_{\tau \sigma}\right) \times G^d_{\tau\sigma}$.}
\begin{equation}
 \det {\mathbb G}_{\tau \sigma} = \left(\prod_{k}g_{k \tau}\right) G^d_{\tau\sigma},  
 \label{determinante}
\end{equation}
and the impurity Dyson equation 
\begin{equation}
 (G^{d}_{\tau\sigma})^{-1} = z - \epsilon_{\tau\sigma} - \Gamma_{\tau}(z) - \Sigma^d_{\tau\sigma}(z), 
\end{equation}
where $\Sigma^d_{\tau\sigma}$ is the impurity self-energy due to the interaction terms (\ref{hint}), we obtain
\begin{equation}
 \frac{\partial \ln \det{\bm {\mathbb G}}^{-1}_{\tau\sigma}}{\partial z} =
 \sum_{k} \frac{\partial \ln g^{-1}_{k \tau}}{\partial z} + \frac{\partial \ln (G^{d}_{\tau\sigma})^{-1}}{\partial z} =
 \sum_{k}g_{k \tau} + \left(1-\frac{\partial \Gamma_{\tau}}{\partial z} 
 -\frac{\partial \Sigma^d_{\tau\sigma}}{\partial z}\right)G^d_{\tau\sigma}. 
 \label{derivada}
\end{equation}
Therefore, 
\begin{equation}
  {\rm Tr}\;{\mathbb G}_{\tau\sigma} = \frac{\partial \ln \det {\mathbb G}^{-1}_{\tau\sigma}}{\partial z} + G^d_{\tau\sigma}(z) \frac{\partial\Sigma^d_{\tau \sigma}(z)}{\partial z}, 
\end{equation}
and the electron numbers are given by 
\begin{equation}
 \langle {N}_{\tau\sigma}\rangle = \oint_{\Gamma} \frac{dz}{2\pi i} n_F(z) 
 \frac{\partial \ln \det {\mathbb G}^{-1}_{\tau\sigma}}{\partial z} + \oint_{\Gamma} \frac{dz}{2\pi i} n_F(z) G^d_{\tau\sigma}(z) \frac{\partial\Sigma^d_{\tau \sigma}(z)}{\partial z}.
 \label{nas}
 \end{equation}

\subsection*{Generalized Friedel sum rules and Luttinger integrals}
 Here, we generalize the Friedel sum rules for multiorbital models using the conservation laws studied by Yoshimori and Zawadowski~\cite{yoshimori82}. 
The first integral in the Supplementary Eq. (\ref{nas}) for the electron numbers can be rewritten, using (\ref{derivada}), as 
\begin{equation}
\sum_{k}  \oint_{\Gamma} \frac{dz}{2\pi i} n_F(z)  g_{k\tau}(z) + \oint_{\Gamma} \frac{dz}{2\pi i} n_F(z) 
\frac{\partial \ln (G^{d}_{\tau\sigma})^{-1}}{\partial z} \equiv \langle N_{\rm \tau \sigma}^{{\rm c}(0)}\rangle 
+\oint_{\Gamma} \frac{dz}{2\pi i} n_F(z)   \frac{\partial \ln (G^{d}_{\tau\sigma})^{-1}}{\partial z}, 
\label{nc0}
\end{equation}
where $\langle N_{\rm \tau \sigma}^{{\rm c}(0)}\rangle$ are the conduction electron numbers in the absence of the
impurity. In the wide-band limit, the number of conduction electrons remains unaltered when the impurity is
introduced. As a consequence  
$\langle N_{\rm \tau \sigma}\rangle-\langle N_{\rm \tau \sigma}^{{\rm c}(0)}\rangle = \langle n_{\tau \sigma}\rangle$. 
On the other hand, the last integral in (\ref{nc0}), in the zero temperature limit,  is
\begin{equation}
  \lim_{T \to 0} \oint_{\Gamma} \frac{dz}{2\pi i} n_F(z)   \frac{\partial \ln (G^{d}_{\tau\sigma})^{-1}}{\partial z} = -\frac{1}{\pi}{\rm Im}\;
  \int_{-\infty}^0  \frac{\partial \ln (G^{d}_{\tau\sigma})^{-1}}{\partial \omega}d\omega = \frac{1}{\pi}{\rm Im}\ln G^d_{\tau\sigma}(0)-\frac{1}{\pi}{\rm Im}\ln G^d_{\tau\sigma}(-\infty) = \frac{\delta_{\tau\sigma}}{\pi}, 
\end{equation}
where we have defined the (Fermi level) phase shift $\delta_{\tau\sigma}$~\cite{lang} through 
$$G^d_{\tau\sigma}(0) = -|G^d_{\tau\sigma}(0)|e^{i\delta_{\tau\sigma}}.$$ 
 Putting all these ingredients together we obtain the generalized Friedel sum rules, that relate the quasiparticle
 phase shifts and the impurity occupancies at zero $T$,
 \begin{equation}
 \delta_{\tau\sigma} = \pi  \langle n_{\tau \sigma}\rangle  + I_{\tau\sigma}, 
 \label{gfsr}
 \end{equation}
with the Luttinger integrals $I_{\tau\sigma}$~\cite{lutti} defined as 
\begin{equation}
 I_{\tau \sigma} = {\rm Im}\int_{-\infty}^0 G^d_{\tau\sigma}\frac{\partial \Sigma^d_{\tau \sigma}}{\partial \omega}d\omega.
 \label{LuttingerIntegral}
\end{equation}
Until very recently, it had been considered that the vanishing of the Luttinger integrals represents one of the
hallmarks of Fermi liquid phases~\cite{lutti,lang}. 
In this way, the (ordinary) Friedel sum rules state that the quasiparticle phase shifts are simply the impurity occupancies divided by $\pi,$   
as it was known from the fifties. 
However, in the last two years, it was shown that, for certain single-impurity~\cite{blesio18,blesio19} and two-impurity~\cite{curtin18,nishi} models, 
the Luttinger integrals can take the discrete values $\pm \pi/2$ in the so-called {\it non-Landau} Fermi liquids (NLFL), that is, Fermi liquid phases 
that are not adiabatically connected with their non-interacting counterparts~\cite{blesio18}.  
In this work we have found that an iron phthalocyanine molecule  on Au(111)  behaves as a NLFL in the absence of an applied magnetic field, while for $B \neq 0$ 
it is an ``ordinary'' Fermi liquid, but with non-zero Luttinger integrals that change continuosly with $B$. 
So, we have shown that there are Fermi liquids in which the Friedel sum rules (\ref{gfsr}) holds only in its
generalized version with non-zero (continuous)
Luttinger integrals.  

As the impurity spectral function at the Fermi level is 
\begin{equation}
 \rho_{\tau\sigma}(0) = -\frac{1}{\pi}{\rm Im}G^d_{\tau\sigma}(0) = \frac{1}{\pi}|G^d_{\tau\sigma}(0)| \sin(\delta_{\tau\sigma}), 
\end{equation}
for a Fermi liquid whose low (but {\it finite}) energy behavior is characterized by ${\rm Im}\Sigma^d_{\tau\sigma}
\propto -a \omega^2$ ($a > 0$). This results in
$|G^d_{\tau\sigma}(\omega\to 0)| = \sin\delta_{\tau\sigma}/\Delta_\tau$. Consequently, the Friedel sum rule for spectral functions, valid for Fermi liquids, 
is given by 
\begin{equation}
 \rho_{\tau \sigma}(0) = \frac{1}{\pi \Delta_\tau} \sin^2\delta_{\tau\sigma}.
 \label{gfsrsf}
\end{equation}
The above argument remains valid if the imaginary part of the self-energy has a Dirac $\delta$ centered exactly at the Fermi level, as it happens in the NLFL~\cite{blesio18}
where ${\rm Im}\Sigma^d_{\tau\sigma} = -a\omega^2 - b\delta(\omega)$. The generalized Friedel sum rule for the spectral funcion $\rho_{\tau\sigma}(\omega\to 0)$ 
(\ref{gfsrsf}) therefore holds even for the NLFL.

\subsection*{Topological interpretation of the Luttinger integrals}

Here, we demonstrate that for certain Anderson impurity models (single-orbital, degenerate two-orbital with an arbitrary magnetic field, non-equivalent two-orbital model without 
magnetic field), the Luttinger integrals have a topological nature. 
For this purpose, we turn off the interactions of the Anderson Hamiltonian $H_A$. 
The total electron numbers can change as the interactions are turned off. Using (\ref{nas}) 
the electron numbers in the non-interacting case are
\begin{equation}
 \langle {N}_{\tau\sigma}\rangle^0 = \oint_{\Gamma} \frac{dz}{2\pi i} n_F(z) 
 \frac{\partial \ln \det ({\mathbb G^{(0)}}_{\tau\sigma})^{-1}}{\partial z}, 
 \label{nas0}
 \end{equation}
where ${\mathbb G^{(0)}}_{\tau\sigma}$ are the non-interacting GF block matrices. 
One should not confound these electron numbers, corresponding to the non-interacting Anderson model ($H=H_0$, i.e.
setting $H_\mathrm{int}=0$) with the impurity still hybridized with the conduction channels, with the 
 previously defined $\langle N_{\tau \sigma}^{{\rm c}(0)}\rangle$
 (Supplementary Eq.~\ref{nc0}) that correspond to the conduction
 electron numbers for the case of decoupled impurity (i.e., setting $H_{\rm hyb} =0$).
 
From (\ref{nas}) and (\ref{nas0}), we obtain 
\begin{equation}
 \langle {N}_{\tau\sigma}\rangle - \langle {N}_{\tau\sigma}\rangle^0  = \oint_{\Gamma} \frac{dz}{2\pi i} n_F(z) 
 \frac{\partial \ln D_{\tau\sigma}}{\partial z} + \oint_{\Gamma} \frac{dz}{2\pi i} n_F(z) G^d_{\tau\sigma}(z) \frac{\partial\Sigma^d_{\tau\sigma}(z)}{\partial z}, 
 \label{deltan}
 \end{equation}
where we have defined the scalar function
\begin{equation}
 D_{\tau\sigma}(z) = \frac{\det {\mathbb G^{(0)}}_{\tau\sigma} }{\det  {\mathbb G}_{\tau\sigma} }. 
\end{equation}
Taking into account (\ref{determinante}), its expression is simplied to
\begin{equation}
 D_{\tau\sigma}(z) = \frac{G^{d(0)}_{\tau\sigma}(z)}{G^d_{\tau\sigma}(z)}.
\end{equation}
For $T \to 0$, it can be shown~\cite{seki17} that the integral 
\begin{equation}
 n_D({\Gamma}) \equiv \oint_{\Gamma} \frac{dz}{2\pi i} n_F(z)  \frac{\partial \ln D_{\tau\sigma}}{\partial z}
\end{equation}
is a winding number. When $T\to 0$,  the vanishing of the Fermi function for ${\rm Re}\; z > 0$  turns ${\Gamma}$ into
the curve ${\Gamma}_{-} \cup {\Gamma}_0$ that encircle only the
negative real axis and the origin (see Supplementary Fig.~\ref{contours}), and
\begin{eqnarray}
 n_D({\Gamma}) & \to &  n_D(\Gamma_-) + \frac{1}{2}n_D(\Gamma_0) = \nonumber\\
 & = &  \oint_{\Gamma_{-}} \frac{dz}{2\pi i} \frac{\partial \ln D_{\tau\sigma}(z)}{\partial z}
 +  \frac{1}{2}\oint_{{\Gamma_0}} \frac{dz}{2\pi i} \frac{\partial \ln D_{\tau\sigma}(z)}{\partial z} =
 \label{winding}
 \\
 &= & \frac{1}{2\pi i} \oint_{{\cal D}_{\Gamma_-}} \frac{dD_{\tau\sigma}}{D_{\tau\sigma}} +
 \frac{1}{2}\times \frac{1}{2\pi i} \oint_{{\cal D}_{\Gamma_0}} \frac{dD_{\tau\sigma}}{D_{\tau\sigma}},\nonumber
 \end{eqnarray}
where ${\cal D}_{\Gamma_-}$ (${\cal D}_{\Gamma_0}$) is the closed curve that describes $D(z),$ as $z$ goes along ${\Gamma}_{-}$ ($\Gamma_0$) 
in the complex plane $({\rm Re}\;D, {\rm Im}\;D)$.
 $n_{D}({\Gamma_-}), n_D(\Gamma_0)$ are the winding numbers of $D$ around the origin for paths $\Gamma_-$ and
 $\Gamma_0$, respectively. 
 These numbers are integer (positive if $D$ winds counterclockwise around the origin, negative in the other direction). The $1/2$ factor that multiply $n_D(\Gamma_0)$ 
 comes from $n_F(z=0)=\frac{1}{2}$.
 
In the  $T\to 0$ limit we can use the relations~\cite{mahan}
\begin{equation}
 \lim_{T \to 0} \oint_{\Gamma} \frac{dz}{2\pi i} n_F(z) F(z) = -\frac{1}{\pi}{\rm Im}\; \int_{-\infty}^0  F(\omega)d\omega 
\end{equation}
to get, from (\ref{deltan}),   
\begin{equation}
 \langle {N}_{\tau\sigma}\rangle - \langle {N}_{\tau\sigma}\rangle^0  = -\frac{1}{\pi}{\rm Im}\;\int_{-\infty}^0 
 \frac{\partial \ln D_{\tau\sigma}}{\partial \omega}d\omega - \frac{1}{\pi}I_{\tau\sigma}. 
 \end{equation}

The integral can be expressed in terms of winding numbers (\ref{winding}), that is, 
\begin{equation}
 \langle {N}_{\tau\sigma}\rangle - \langle {N}_{\tau\sigma}\rangle^0  = n_D(\Gamma_-) + \frac{1}{2}n_D(\Gamma_0) - \frac{1}{\pi}I_{\tau\sigma}.  
 \end{equation}
From this we deduce that if the total electron numbers do not change when the interactions are turned off, the
Luttinger integrals only take discrete values that are multiples of $\pi/2$:
\begin{equation}
 I_{\tau\sigma} = \pi n_D(\Gamma_-) + \frac{\pi}{2}n_D(\Gamma_0).
\end{equation}
 Note that  $n_D(\Gamma_0)$ is different from zero if $D(z)$ has a zero or a pole at $z=0$.  For example, if $D(z)=z$ or $1/z$ and we take $\Gamma_0$ as a circle of radius $R$ centered
 in the origin, then 
 \begin{equation}
  n_D(\Gamma_0) = \frac{1}{2\pi i}\lim_{R\to 0} \oint_{\Gamma_0(R)} \frac{dD}{D} =  \frac{1}{2\pi i}\int_{0}^{2\pi} \frac{(\pm i) R e^{\pm i\theta} d\theta}{R e^{\pm i\theta} } = \pm 1.
 \end{equation}

\subsection*{Effective one-body Hamiltonian}

If, as it can in general be expected, $\langle {N}_{\tau\sigma}\rangle \neq \langle {N}_{\tau\sigma}\rangle^0,$ we can
try to tune the effective impurity one-body energies (chemical potential, magnetic field, 
crystal-field) in order to force the same interacting and non-interacting electron numbers. If this is possible, we will show that the Luttinger integrals keep their topological nature. 
We define an effective one-body impurity Hamiltonian $\tilde{H}_d$
\begin{equation}
 \tilde{H}_d = \sum_{\tau\sigma} \left(\tilde{\epsilon} - \frac{\sigma}{2}\tilde{B} + \tau \tilde{C}\right) n_{\tau\sigma} 
 \equiv \sum_{\tau\sigma}\tilde{\epsilon}_{\tau\sigma}{n}_{\tau\sigma}.
 \label{heff}
\end{equation}
The dependence of the single-particle energy $\tilde{\epsilon}_{\tau \sigma}$ on $\sigma$ and $\tau$ is dictated by the requirement that the effective Hamiltonian obey 
the same conservation laws (see Supplementary Note 1) as the original Anderson Hamiltonian.

 To go further we re-arrange the non-interacting and interacting terms of the impurity Hamiltonian: 
 \begin{equation}
  {H}_{\rm imp} = {H}_d + {H}_{\rm int} = \tilde{H}_d + \tilde{H}_{\rm int}, 
 \end{equation}
with 
\begin{equation}
 \tilde{H}_d = \sum_{\tau\sigma} \tilde{\epsilon}_{\tau\sigma}{n}_{\tau\sigma} = {H}_d + 
  \sum_{\tau\sigma} \left(\tilde{\epsilon}_{\tau\sigma}-\epsilon_{\tau\sigma}\right){n}_{\tau\sigma}, 
\end{equation}
while
\begin{equation}
 \tilde{H}_{\rm int} = {H}_{\rm int} +\sum_{\tau\sigma} \left(\epsilon_{\tau\sigma}-\tilde{\epsilon}_{d\tau\sigma}\right){n}_{\tau\sigma}.
\end{equation}
As the Hamiltonian remains unaltered, the impurity Green function does not change. However, there is a rearrangement of its ``non-interacting'' and ``interacting'' contributions: 
\begin{equation}
(G^d_{\tau\sigma})^{-1} = (G^{d(0)}_{\tau\sigma})^{-1} - \Sigma^d_{\tau\sigma} = z - \epsilon_{\tau\sigma} - \Gamma_{\tau} - \Sigma^d_{\tau\sigma} = 
\end{equation}
\begin{equation}
=z -\tilde{\epsilon}_{\tau\sigma} - \Gamma_{\tau} - \Sigma^d_{\tau\sigma} + (\tilde{\epsilon}_{\tau\sigma}-{\epsilon}_{\tau\sigma}) =  
(\tilde{G}^{d(0)}_{\tau\sigma})^{-1}- \tilde{\Sigma}^d_{\tau\sigma}, 
\end{equation}
where the renormalized self-energy is defined as the original one shifted by a constant real number (a Hartree contribution to the self-energy):
\begin{equation}
\tilde{\Sigma}^d_{\tau\sigma} = \Sigma^d_{\tau\sigma} +   (\tilde{\epsilon}_{\tau\sigma}-{\epsilon}_{\tau\sigma}).
\end{equation}
We can see that the renormalization does not change the Luttinger integrals  
\begin{equation}
 \tilde{I}_{\tau\sigma} = {\rm Im}\;\int_{-\infty}^0  G^d_{\tau\sigma} \frac{\partial \tilde{\Sigma}^d_{\tau\sigma}}{\partial \omega} d\omega= I_{\tau\sigma}, 
\end{equation}
as the constant shift of the self-energy is irrelevant due to the derivative.

Therefore, if we are able to tune the effective parameters so that all the ``effective non-interacting'' electron numbers coincide with the interacting ones, that is
\begin{equation}
 \langle {N}_{\tau\sigma}\rangle = {\langle \tilde{N}_{\tau\sigma}\rangle} ^{0}, 
 \label{condition}
 \end{equation}
then, the topological relation between the Luttinger integrals and the winding numbers 
\begin{equation}
 I_{\tau\sigma} = -{\rm Im}\;\int^0_{-\infty} \frac{\partial \ln \tilde{D}_{\tau\sigma}}{\partial \omega}d\omega = \pi n_{\tilde{D}}(\Gamma_-) + 
 \frac{\pi}{2}n_{\tilde D}(\Gamma_0),
\end{equation}
is still valid, with 
\begin{equation}
 \tilde{D}_{\tau\sigma} = \frac{\tilde{G}^{d(0)}_{\tau\sigma}}{G^d_{\tau\sigma}}.
\end{equation}

The possibility to satisfy the electron number conditions $\langle {N}_{\tau\sigma}\rangle = \langle
{\tilde N}_{\tau\sigma}\rangle^{0}$ for an Anderson Hamiltonian by tuning the effective non-interacting parameters, depends in how many conditions we must enforce, and how many effective parameters we have at our disposal for this purpose.   
For a single orbital Anderson model under a magnetic field, there are only two electron conduction numbers conditions,
i.e. Summplementary Eq.~(\ref{condition}) for $\sigma=\uparrow$ and $\sigma=\downarrow$, to satisfy and we can use two effective parameters
($\tilde{\epsilon}, \tilde{B}$). Therefore, the conditions can be satisfied and the Luttinger integrals have a topological nature. 
For a degenerate two-orbital Anderson model, even under a magnetic field, $\langle N_{1\sigma}\rangle = \langle
N_{-1\sigma}\rangle$ always holds due to the degeneracy. 
As a consequence, there are again two electron conduction numbers conditions, and both can be satisfied by tuning the effective
parameters. Finally, for the Anderson model corresponding to FePc on Au(111), there are two non-equivalent orbitals. If $B=0$, the spin symmetry $\langle N_{\tau\uparrow}\rangle = 
\langle N_{\tau\downarrow}\rangle$ reduces the electron number conditions to two, and, hence, they can be satisfied. However, in the presence of a finite magnetic field, 
there are four different electron number conditions $\langle {N}_{\tau\sigma}\rangle = \langle {\tilde{N}}_{\tau\sigma}\rangle^{0}$ with only three parameters at our disposal 
($\tilde{\epsilon}, \tilde{B}, \tilde{C}$). Therefore, it would be impossible to make this fit for all the electron numbers. In this last case, the individual 
Luttinger integrals may not have a topological (discrete) nature. In fact, numerically we have found that $I_{\tau\sigma}$ change continuosly with $B$, and that this behavior is essential to understand the 
experimental STM spectra of FePc on Au(111) under a magnetic field.

Even though each individual Luttinger integral loses its topological nature for the non-equivalent two-channel Anderson Hamiltonian (\ref{hamilton}), in the next section, we will demonstrate 
that, due to the conservation laws, certain linear combinations of Luttinger integrals do have a topological character. 

\subsection*{Conservation laws and topological numbers}

As already mentioned, for the non-degenerate two-orbital Anderson model, 
it is impossible to match the interacting and non-interacting electron number for each spin-orbital separately. However, with the help of the {\bf three} effective parameters: 
chemical potencial $\tilde{\epsilon}$, magnetic field $\tilde{B},$ and crystal field $\tilde{C}$, using the non-interacting GF's  corresponding to the 
effective non-interacting Anderson Hamiltonian 
$$\tilde{G}_{\tau \sigma}^{d(0)} = \frac{1}{\omega + i0^+ - \tilde{\epsilon} -\sigma \frac{\tilde{B}}{2} + \tau \tilde{C}-\Gamma_\tau(\omega)},
$$
we can match, between the interacting and non-interacting models, the {\bf three} conserved quantities (see Supplementary Note 1): \\

\noindent
i) the total electron number  $N$ 
\begin{equation}
N -  \sum_{\tau\sigma}\langle N^{{\rm c}(0)}_{\tau\sigma}\rangle \equiv  \sum_{\tau\sigma} \langle n_{\tau\sigma}\rangle = \sum_{\tau\sigma} 
\langle \tilde{n}_{\tau\sigma}\rangle ^{0} = -\frac{1}{\pi}\sum_{\tau\sigma} \arctan\left(\frac{\Delta_\tau}{\tilde{\epsilon}-\sigma\frac{\tilde{B}}{2}+ \tau\tilde{C}}\right).
\end{equation}
ii) the total $S^{\rm tot}_z$  
\begin{equation}
 S_z^{\rm tot} = \frac{1}{2}\sum_{\tau\sigma} \sigma \langle n_{\tau\sigma}\rangle = 
 \frac{1}{2}\sum_{\tau\sigma} \sigma \langle \tilde{n}_{ \tau\sigma}\rangle^{0} 
 = -\frac{1}{2\pi}\sum_{\tau\sigma} \sigma\arctan\left(\frac{\Delta_\tau}{\tilde{\epsilon}-\sigma\frac{\tilde{B}}{2}+ \tau\tilde{C}}\right),
 \end{equation}
and, iii) the total orbital isospin $T_z$ 
\begin{equation}
T_z \equiv N_1  - N_{-1} \equiv \sum_{\tau\sigma} \tau\langle n_{\tau\sigma}\rangle = 
\sum_{\tau\sigma} \tau\langle \tilde{n}_{\tau\sigma}\rangle ^{0} 
= -\frac{1}{\pi}\sum_{\tau\sigma} \tau\arctan\left(\frac{\Delta_\tau}{\tilde{\epsilon}-\sigma\frac{\tilde{B}}{2}+ \tau\tilde{C}}\right).
  \end{equation}

In this way, we can show that the following linear combination of Luttinger integrals 
\begin{eqnarray}
 T  & \equiv & \sum_{\tau\sigma} I_{\tau\sigma}   =  -\sum_{\tau\sigma} {\rm Im}\;\int^0_{-\infty}\frac{\partial \ln \tilde{D}_{\tau\sigma}}{\partial \omega}d\omega, \\
 T_\sigma & \equiv & \sum_{\tau\sigma} \sigma I_{\tau\sigma}   =  -\sum_{\tau\sigma}\sigma\; {\rm Im}\;\int^0_{-\infty}\frac{\partial \ln \tilde{D}_{\tau\sigma}}{\partial \omega}d\omega, 
 \label{ts}
 \\
 T_\tau & \equiv & \sum_{\tau\sigma} \tau I_{\tau\sigma}   =  -\sum_{\tau\sigma} \tau \;{\rm Im}\;\int^0_{-\infty}\frac{\partial \ln \tilde{D}_{\tau\sigma}}{\partial \omega}d\omega, \\
\end{eqnarray}
are topological numbers, as the left hand integrals can be expressed in terms of winding numbers. By the numerical NRG
computation of the Luttinger integrals, we have found that, for the model describing FePc on Au(111), 
$T_\sigma = T_\tau = 0$. Meanwhile $T$ takes the non-trivial topological value $2\pi$ for $D>D_c$ and $B=0$ and goes to zero when the magnetic field is turned on, signalling a topological transition from a NLFL to 
an OFL (with non-zero Luttinger integral) as a function of the applied magnetic field.

\section*{Supplementary Note 3: Magnetic field dependence of the impurity self-energy}
Taking into account that $T_\tau = T_\sigma = 0$, the Luttinger integrals can be expressed as 
\begin{equation}
 I_{1\uparrow} = I_{-1\downarrow} = I_0 - \alpha(D,B), \;\; I_{1\downarrow} = I_{-1\uparrow} = I_0 + \alpha(D,B), 
\end{equation}
where $I_0 = T/4$ has a topological (discrete) nature, while $\alpha(D,B)$ is the non-topological component of the Luttinger integrals.  For $B=0$  ($D > D_c$)  
all the Luttinger integrals are equal, $I_{\tau \sigma} = \pi/2$. When a small positive $B$ is turned on, $I_{-1\uparrow} (=I_{1\downarrow})$ changes continuosly from $\pi/2$ 
while $I_{1\uparrow} (=I_{-1\downarrow})$ has a jump to $ -\pi/2$. For $B < 0$ these behaviors are reversed. 
The abrupt change in the Luttinger integrals can be ascribed to the shift of the $B=0$ Fermi level  pole of the imaginary part of the self-energies: for 
$\Sigma^d_{1\uparrow}, \Sigma^d_{-1\downarrow}$ a small positive $B$ shifts (and broadens) the pole to negative energies of the order of $-B/2$, giving rise to a $\pi$ change in the 
Luttinger integral. 
On the other hand, for $\Sigma^d_{1\downarrow}, \Sigma^d_{-1\uparrow}$, due to the symmetry ${\rm Im}\Sigma^d_{\tau\sigma}(\omega) = {\rm Im}\Sigma^d_{\tau -\sigma}(-\omega)$, 
the poles go to positive energies.  These behaviors for ${\rm Im}\Sigma^d_{1\uparrow}, {\rm Im}\Sigma^d_{-1\downarrow}$ are 
displayed in Supplementary Fig.~\ref{selfenergies}.  

\begin{figure} [ht] 
\begin{center}
\includegraphics*[width=0.47\columnwidth]{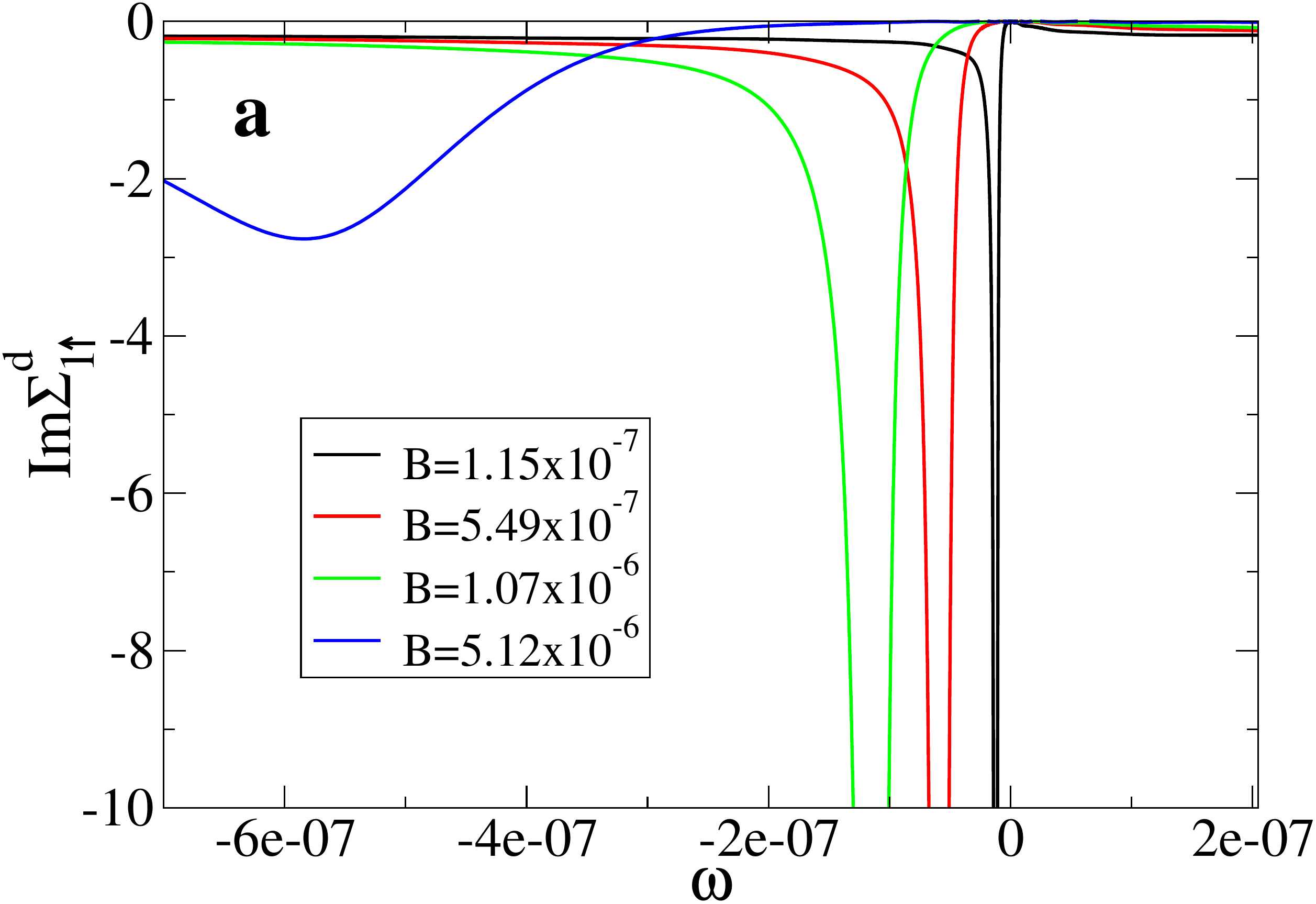} 
\hspace*{0.11cm} 
\includegraphics*[width=0.47\columnwidth]{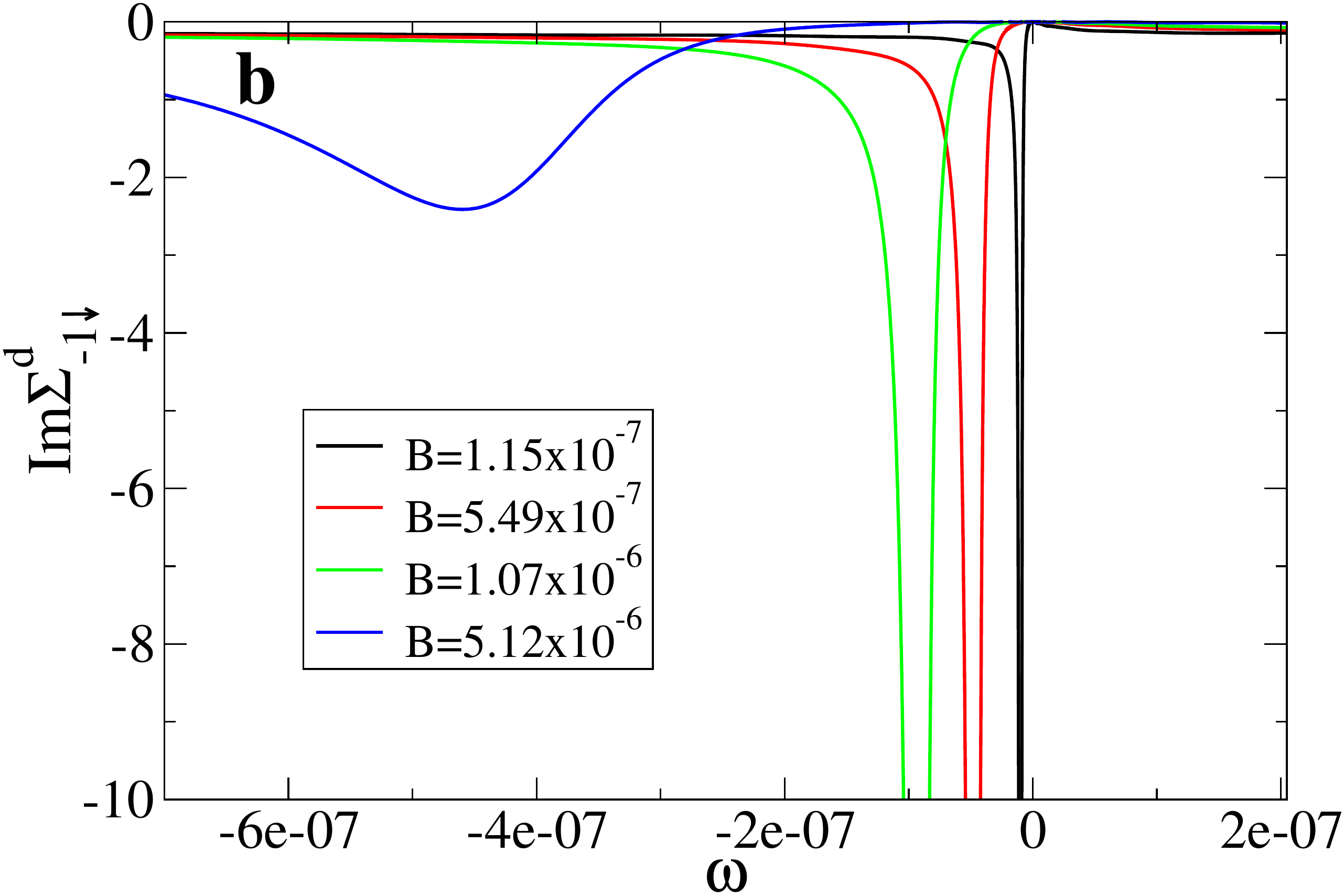} \\
\caption{{\bf Imaginary part of the self-energies.} \textbf{a)}  $\Sigma^d_{1\uparrow}$ and \textbf{b)} $\Sigma^d_{-1\downarrow}$, 
close to the Fermi level, for different magnetic fields. The Anderson Hamiltonian parameters are the same as in the main text, except for $D=1.08 D_c$.}
\label{selfenergies}
\end{center}
\end{figure}

\section*{Supplementary Note 4: Detailed comparison with experiment}  
  
In this note, we compare side by side experimental results and our theory 
for three sets of experiments.

\subsection*{Dependence of the differential conductance as the molecule is
separated from the surface}

\begin{figure} [ht] 
\begin{center}
\includegraphics*[width=0.80\columnwidth]{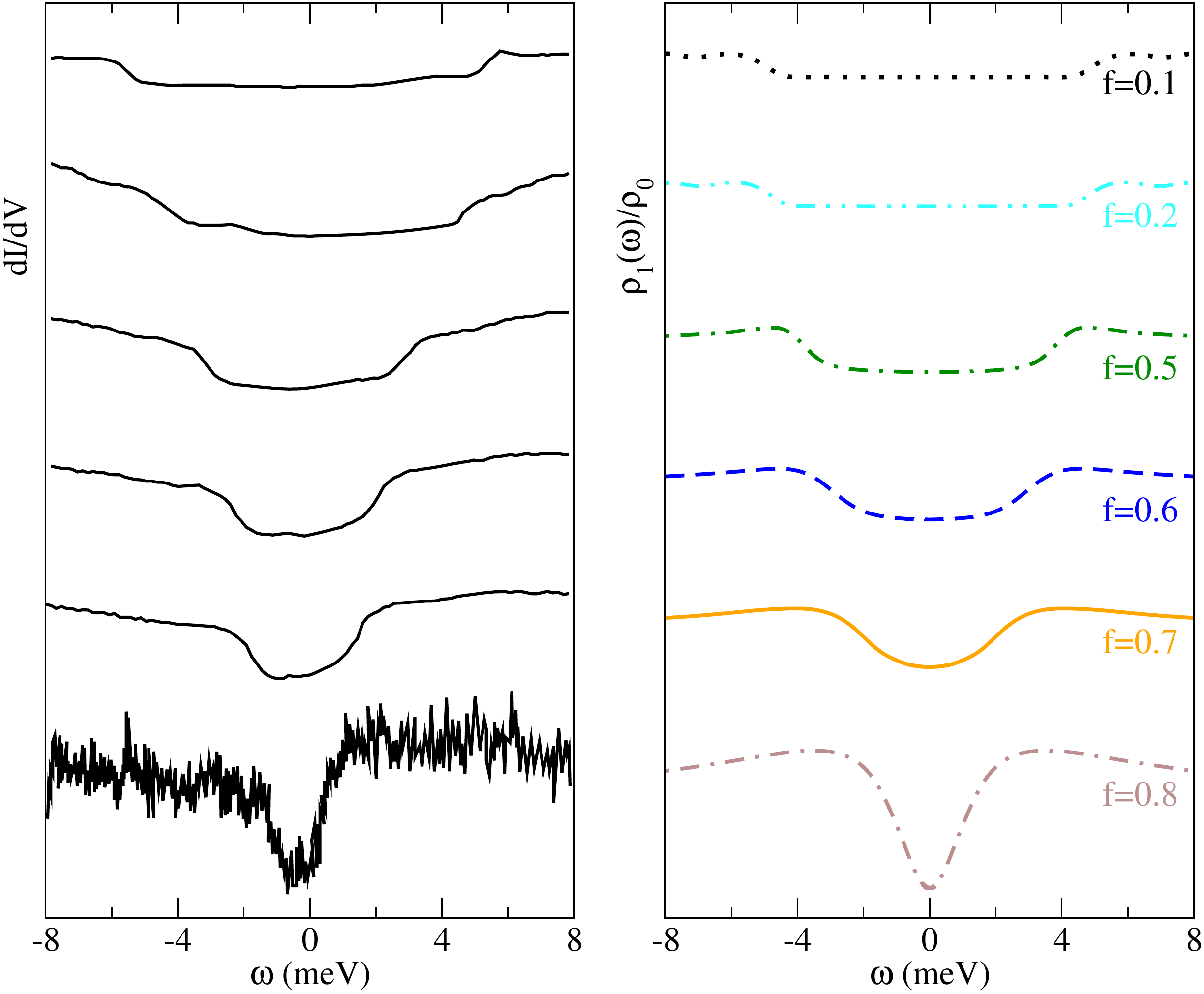} 
\caption{{\bf Raising molecule from surface.}
Left: experimental differential conductance as a function of voltage
(taken from Fig. 2 (b) of Ref. \cite{hira}). Right: theoretical spectral density of level 1 as a function of energy, as the molecule is raised from the surface. The spectral densities plotted are the same as in Fig. 2 of the main text, where a rescaling along the vertical axis was made for a direct comparison with experimental data.}
\label{fhira}
\end{center}
\end{figure}

In Supplementary Fig.~\ref{fhira} we represent a set of experimental curves 
for the differential conductance $dI/dV$ taken from
Ref. \cite{hira} as the molecule is raised from the surface and compare
them with the corresponding spectral density of the localized
states of symmetry $3z^2-r^2$ in our model. Since the experimental 
results are affected by a factor that decays exponentially with the 
distance between the tip and the surface, which is not included in our
model, we have multiplied them by an arbitrary factor $f$.

In general, the shapes of experimental and theoretical curves are very similar, except for large separations (top) for which the theoretical
curves are affected by the resolution of the NRG calculations
and for small separations (bottom) for which some admixture 
of conduction states is expected to influence the conductance
and is taken into account by the factor $q$ in Eq. (8) of the main
text. In particular, the comparison when the molecule is on the surface
is discussed below including this factor.

\subsection*{Differential conductance for different temperatures}

\begin{figure} [ht] 
\begin{center}
\includegraphics*[width=0.80\columnwidth]{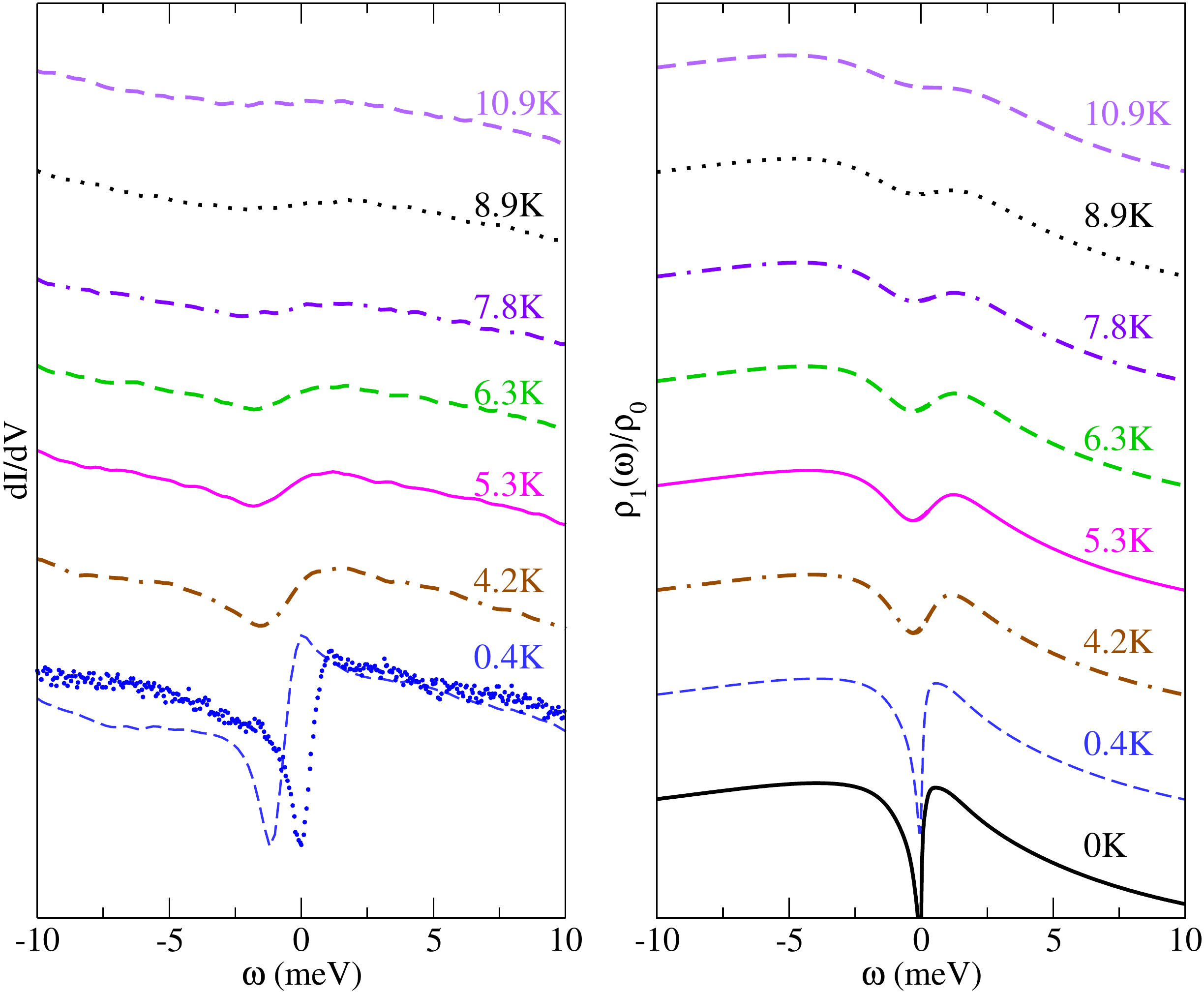}
\caption{{\bf Temperature dependence.}
Left: experimental differential conductance as a function of voltage
for several temperatures. Curves with different type of lines are
taken from Fig. 2 (b) of Ref. \cite{yang} and data in dots
($T=0.4\,K$) is taken with permission from Fig. 1 (c) of Ref. \cite{mina}. Right: temperature dependence of the differential conductance as a function of voltage as in Fig. 3 of the main text with a vertical displacement for a better comparison with the left panel.}
\label{tsup}
\end{center}
\end{figure}

In Supplementary Fig.~\ref{tsup} we compare experiment (taken mainly from Ref. \cite{yang}) 
and theory for the differential conductance $dI/dV$ at different 
temperatures. The trend of experimental and theoretical curves
is very similar. The theoretical curves have a steeper downwards
slope for positive voltage, and in the experiment the maximum in $dI/dV$
for positive voltage is larger than the relative maximum for negative 
voltage. Both features might be related with the effect of 
subtraction of a background or other features of a particular experiment.  
At the lowest experimental temperature (0.4 K) we also show
the experimental curve taken from Fig. 1 (c) of Ref. \cite{mina}.
This allows the reader to grasp some experimental uncertainties.

\subsection*{Differential conductance for different magnetic fields}

\begin{figure} [ht] 
\begin{center}
\includegraphics*[width=0.80\columnwidth]{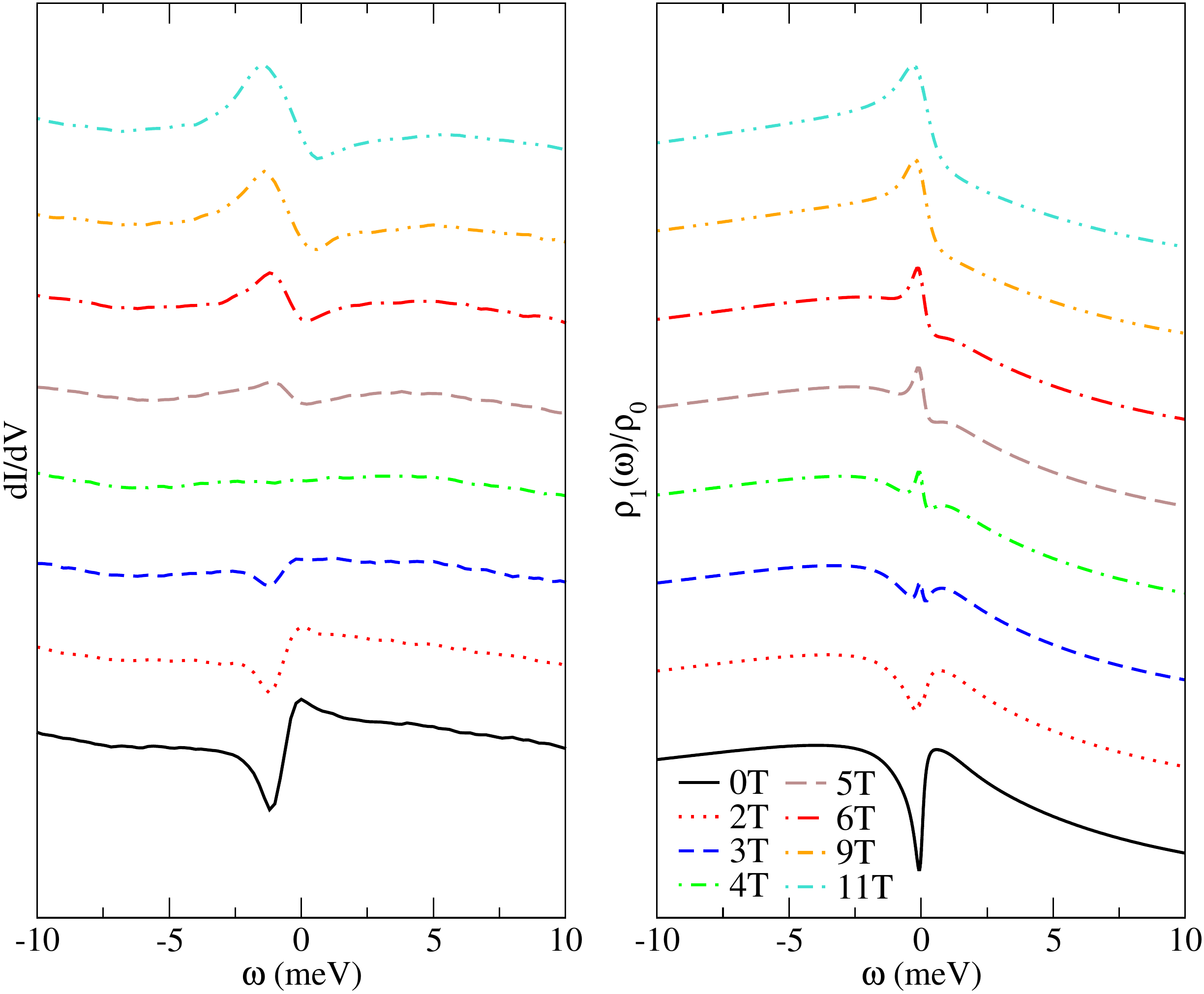}
\caption{{\bf Magnetic field dependence.}
Left: experimental differential
conductance as a function of voltage for several values of the
magnetic field (taken from Fig. 2 (c) of Ref. \cite{yang}). Right: magnetic field dependence of the differential conductance as a function of voltage as in Fig. 4 of the main text with a vertical displacement for a better comparison with the left panel.}
\label{bsup}
\end{center}
\end{figure}

In Supplementary Fig.~\ref{bsup} we compare the experimental results for $dI/dV$ (also taken from Ref. \cite{yang}) and theory for different magnetic fields, 
assuming a gyromagnetic factor 
$g=2$, inside the range of uncertainty of values reported in the  
supplemental material of Refs. \cite{hira}, \cite{yang}. Taking into account
the features mentioned above related with experimental uncertainties, and the absence of alternative physical explanations,
the agreement is very good.

\end{widetext}


\begin{thebibliography}{99}
\bibitem{note} While three channels participate in FePc \cite{joaq} 
including all of them in an NRG calculation would
lead to a too fast increase in the Hilbert space as a function of the iteration rendering
impossible to obtain precise results.
\end{thebibliography}
\end{document}